\documentclass[]{aa}
\usepackage{natbib}
\usepackage{graphicx}
\usepackage{txfonts}

\newcommand{\Lya}{\mbox{Ly $\alpha$\,}}
\newcommand{\logn}{\mbox{${\rm log}~N_{\rm \ion{H}{i}}$}}
\newcommand{\kms}{\mbox{km\,s$^{-1}$}}
\newcommand{\N}{\mbox{column density}}
\newcommand{\cddf}{\mbox{column density distribution function}}
\newcommand{\Ns}{\mbox{column densities}}
\newcommand{\B}{\mbox{Doppler parameter}}
\newcommand{\z}{\mbox{redshift}}

\begin{document}

\title
{The Evolution of Lyman $\alpha$ Absorbers 
in the Redshift Range $0.5<z<1.9$}

\author
{E. Janknecht\inst{1}, D. Reimers\inst{1}, S. Lopez\inst{2} 
and D. Tytler\inst{3}}

\authorrunning{E. Janknecht et al.}
\titlerunning{The Evolution of Lyman $\alpha$ Absorbers at $0.5<z<1.9$}

\offprints{E. Janknecht,\\
e-mail: ejanknecht@hs.uni-hamburg.de}
\institute{Hamburger Sternwarte, Universit\"at Hamburg, Gojenbergsweg 112,
D-21029 Hamburg, Germany\\
\email{[ejanknecht,dreimers]@hs.uni-hamburg.de}
\and
Departamento de Astronomia, Universidad de Chile, Casilla 36-D, Santiago, 
Chile\\
\email{slopez@das.uchile.cl}
\and
Center for Astrophysics and Space Sciences, University of California, 
San Diego, MS 0424, La Jolla, CA 92093-0424, USA
\email{tytler@ucsd.edu}}

\date{Received \today / Accepted}

\abstract{
We investigate the evolution and the statistical properties 
of the {\Lya} absorbers of the intergalactic medium (IGM) 
in the largely unexplored redshift range $z=0.5-1.9$.
We use high-resolution ($R \geq 30\,000$) UV (STIS) and optical (VLT/UVES
and Keck/HIRES) spectra of nine bright quasars with $z_{\rm em} < 1.94$. 
The {\Lya} lines detected in the lines of sight (LOS) towards these quasars
are evaluated with a software package which determines simultaneously
the quasar continuum and the line profiles.
The main results for the combined {\Lya} line sample are summarized 
as follows:

\begin{enumerate}

\item{The evolution of the number density of the absorbers can be described 
by the power law $\frac{{\rm d}n}{{\rm d}z} \propto (1+z)^{\gamma}$.
The number density of the low column density lines 
($N_{\rm \ion{H}{i}}=(10^{12.90}-10^{14.00})$ cm$^{-2}$)
decreases with decreasing $z$ with $\gamma=0.74\pm0.31$ in the interval 
$z=0.7-1.9$. \,\, A comparison with results at higher redshifts shows 
that it is decelerated in the explored redshift range 
and turns into a flat evolution for $z \rightarrow 0$.
The stronger absorbers (\mbox{$N_{\rm \ion{H}{i}} > 10^{13.64}$ cm$^{-2}$}) 
thin out faster ($\gamma=1.50\pm0.45$). 
The break in their evolution predicted for $z=1.5-1.7$ cannot be seen 
down to $z=0.7$. On the other hand, a comparison with values 
from the literature for the local number density gives a hint that this break
occurs at lower redshift.}

\item{The distribution of the column densities of the absorbers is complete 
down to \mbox{$N_{\rm HI}=10^{12.90}$ cm$^{-2}$}. It can be approximated 
by a single power law with the exponent $\beta=1.60\pm0.03$ over almost 
three orders of magnitude. $\beta$ is redshift independent.}

\item{The {\Lya} lines with lower column densities 
as well as the higher column density lines show marginal clustering 
with a $2\sigma$ significance over short distances ($\Delta v < 200$ {\kms} 
and $\Delta v < 100$ {\kms}, respectively). We do not see any difference 
in the clustering with either column density or redshift.}

\item{The distribution of the Doppler parameters has a mean value 
of $\overline{b}=(34\pm22)$ {\kms}. This value is typical 
for the analyzed region. It does not change significantly with $z$.}

\end{enumerate}

}

\maketitle


\section{Introduction}

It is traditional to decompose spectra of the {\Lya} forest into
individual absorbers that can be fit with Voigt profiles, 
each with a redshift $z$, a column density of neutral hydrogen 
$N_{\rm \ion{H}{i}}$ and a Doppler parameter $b$.
Examining a sufficiently large number of LOS 
to background quasi-stellar objects (QSOs), 
one can derive the distributions of these parameters 
as well as dependences between them and in this way deduce characteristics 
of the {\Lya} forest. Observers have often focused on the evolution 
of the number density of {\Lya} lines, usually approximated 
by the power law

\begin{equation}
\label{dn/dz}
\frac{{\rm d}n}{{\rm d}z} = 
\left(\frac{{\rm d}n}{{\rm d}z}\right)_{0} (1+z)^{\gamma}
\end{equation}
\citep{Sargent}.
At high redshift ($z>1.5$), the evolution of the strong lines
(log~$N_{\rm \ion{H}{i}}>14.00$) is steep ($\gamma=2-3$),
transiting into a flat evolution ($\gamma=0.1-0.3$) at lower redshift.
This abrupt slow-down in the evolution, first detected by observers 
\citep{Bahcall, Impey1996, Weymann} and later also seen in simulations
\citep{Theuns1998, Dave1999}, is suspected to happen at \mbox{$z=1.5-1.7$}. 

However, since the apparent break in the evolution rate seems to occur
just at the transition from high-resolution optical spectra to low-resolution
spectra taken with the HST, the location of this break needs to be confirmed
using high-resolution UV spectra. In a previous study, using UV observations
of the extremely bright $z_{\rm em}=1.73$ QSO HE~0515-4414 with a resolution
of 30\,000, it appeared that the slow-down in the evolution rate did not occur
earlier than $z \sim 1$ \citep{Janknecht2002}. However, at low redshift
there are few absorbers, and more LOS are necessary. 

In this paper we use UV and optical high-resolution data of nine bright 
quasars to explore the {\Lya} forest in the redshift range $z=0.5-1.9$.
We believe that this is the first study of the lower column density absorbers 
in the redshift region $z<1.5$ using spectra with $R \geq 30\,000$.
The paper is organized as follows: in Section 2, we explain 
our selection criteria for the quasars, describe the observations 
and define the redshift ranges that we investigate. Section 3 illustrates 
how the {\Lya} lines in the spectra were detected and modelled. 
After mentioning some interesting individual features in the individual LOS 
in Section 4, we discuss our results in Section 5. The main results 
are summarized in Section 6.


\section{Observations}
\subsection{Selection of the quasars} 

Table \ref{quasar parameters} gives an overview of the quasars
whose spectra we selected to analyze the {\Lya} forest. We list their emission 
redshifts $z_{\rm em}$ and their apparent $B$ magnitudes $m_{B}$. 
We selected QSOs with emission redshifts suitable to show the evolution of the
{\Lya} forest at $1<z<2$.
In Table \ref{data} we give the spectral resolution and the approximate $S/N$ 
of the spectral regions that we examined.
The resolution of the spectra of all nine quasars is high enough 
to fully resolve all {\Lya} lines. The $S/N$ varies from one spectrum 
to the next, and within each spectrum, so the minimum observable column density
depends strongly on the spectrum and the wavelength.

\begin{table}
\begin{center}
\begin{tabular}{|l|l|c|}\hline

\hspace{0.7cm} QSO & $z_{\rm em}$ & $m_{B}$ [mag] \\
\hline
PG 1634+706  & 1.34 & 14.9 \\
PKS 0232-04  & 1.44 & 16.6 \\
PG 1630+377  & 1.48 & 16.5 \\
PG 0117+213  & 1.50 & 16.1 \\
HE 0515-4414 & 1.73 & 15.0 \\
HE 0141-3932 & 1.80 & 16.2 \\  
HE 2225-2258 & 1.89 & 16.3 \\
HS 0747+4259 & 1.90 & 15.8 \\
HE 0429-4901 & 1.94 & 16.2 \\
\hline

\end{tabular}
\end{center}
\caption[]
{\label{quasar parameters} Parameters of the quasars}
\end{table}

\subsection{Spectra and data reduction} 

Because the {\Lya} line is shifted into the optical for $z \gtrsim 1.5$,
a complete analysis of the {\Lya} forest at median and high redshifts
requires both optical and UV spectra.
For two of the nine quasars, HE~0515-4414 and HS~0747+4259 
(discovery of both objects published in Reimers et al., 1998; 
for the {\Lya} forest in the LOS to HE~0515-4414 alone see 
Janknecht et al., 2002), data in both spectral regions have been obtained:
in the UV with HST/STIS with the E230M echelle mode ($R \sim 30\,000$), 
and in the optical with UVES ($R \sim 50\,000$) for HE~0515-4414 
and with Keck/HIRES ($R \sim 38\,000$) for HS~0747+4259.

The sample was extended with the UVES spectra of three bright quasars from
the Hamburg ESO Survey taken in the service mode with the VLT. These three
were HE~0141-3932, HE~2225-2258 \citep{Wisotzki} and HE~0429-4901 (not yet
published). The last was taken with a slit width of $d=0.8^{\prime \prime }$, 
the other two with $d=1.0^{\prime \prime }$. The {\Lya} lines 
detected in these spectra strengthen the statistics for $1.5 < z < 2.0$.

Finally, the QSO sample was supplemented in the UV ($0.5<z<1.5$) with data 
from the STIS archive.  For this purpose the spectra of four quasars --- 
PKS~0232-04 \citep{Shimmins}, PG~1630+377 \citep{Noguchi}
and PG~0117+213 \citep{Schmidt} from an observing program of B.~Jannuzi 
(HST proposal 8673/cycle 9) and PG~1634+706 \citep{Schmidt} 
from two programs of S.~Burles (7292/7) and B.~Jannuzi (8312/8) ---
were downloaded from the STIS data base. These data were also taken 
with the E230M echelle mode resulting in the same resolution applicable
to the other STIS spectra.  

Table \ref{data} lists the observing dates, exposure times 
and quality informations for all the QSO spectra.

\begin{table*}
\begin{center}
\begin{tabular}{|l|l|l|r|c|c|}\hline

\hspace{0.7cm} \normalsize{QSO} & \normalsize{Telescope / spectrograph} & 
\normalsize{Observing date} & \normalsize{Exposure time [s]} & 
\normalsize{$R$} & \normalsize{$S/N$ per pixel} \\
\hline
\hline
PG 1634+706  & HST/STIS & May 5, June 26, 1999 & 26\,400-29\,000 & 
30\,000 & 5--50\\ 
\hline
PKS 0232-04  & \footnotesize{HST/STIS} & Feb. 6, 8, 2001, Jan. 19, 2002 
& 41\,900 & 30\,000 & 4--15\\
\hline
PG 1630+377  & \footnotesize{HST/STIS} & Feb. 27, Oct. 8, 12, 2001 & 34\,100 
& 30\,000 & 5--11\\
\hline
PG 0117+213  & \footnotesize{HST/STIS} & Dec. 31, 2000--Jan. 12, 2001 
& 42\,000 & 30\,000 & 4--12\\
\hline
\raisebox{-1ex}{HE 0515-4414} & \footnotesize{HST/STIS} & 
Jan. 31--Feb. 2, 2000 & 31\,500 & 30\,000 & $\sim 10$ \\
& \footnotesize{VLT/UVES} & Oct. 7, 2000--Jan. 3, 2001 
& 31\,500 & 50\,000 & 10--50\\
\hline
HE 0141-3932 & \footnotesize{VLT/UVES} & July 19, Aug. 14--24, 2001 
& 39\,600 & 40\,000& $\sim 25$\\
\hline
HE 2225-2258 & \footnotesize{VLT/UVES} & June 17--28, July 14, 15, 2001 
& 41\,800 & 40\,000 & $\sim 25$ \\
\hline
\raisebox{-1ex}{HS 0747+4259} & \footnotesize{HST/STIS} & Sep. 6, 12--16, 2001 
& 54\,200& 30\,000& 3--10 \\
& \footnotesize{Keck I/HIRES} & Feb. 28, Mar 1, 2001 & 5\,400 & 38\,000 
& 3--12\\
\hline
HE 0429-4901 & \footnotesize{VLT/UVES} & Feb. 1, Mar 18, 19, 2001
& 10\,800& 50\,000 & $\sim 8$\\
\hline
 
\end{tabular}
\caption[]
{\label{data} 
Log of the observations used in this study}
\end{center}
\end{table*}

The data reduction was performed using the pipelines of UVES and STIS
on their respective data.
The Keck data of HS~0747+4259 was reduced by J.~O'Meara
using an internal pipeline and the data reduction software package 
of T.~Barlow. For the HE~0515-4414 STIS spectra another rework had to be made:
the radiative flux was in places smaller than zero because 
an inaccurate correction of the noise background of the CCD chips was used 
in the pipeline. The problem was solved by measuring the signal needed 
to make the minimum flux in saturated lines zero. A slightly reduced background
value (varying from order to order, and extrapolated to orders that lacked 
saturated lines) was then subtracted from the raw data flux, 
resulting in positive flux values for the entire spectrum, 
apart from random photon noise.

The vacuum and the baryocentric corrections to the wavelength scales   
were performed with the data analysis package MIDAS.
Multiple exposures of each QSO were added, weighting each flux value 
$F_{\lambda, i}$ by the inverse variance (noise) $\sigma_{i}^{-2}$.

\begin{table*} 
\begin{center}
\begin{tabular}{|l|c|c|c|l|r|r|r|}\hline 

\hspace{0.5cm} QSO & $\lambda\lambda$ [\AA] & $z$ range & $\Delta\, z$ 
& $\Delta\, X$ & $n$ \\ 
\hline 
\hline 
PG 1634+706  & $1865-2790$ & $0.534-1.295$ & 0.761 & 1.665 & 195 \\ 
\hline 
PKS 0232-04  & $2280-2941$ & $0.876-1.419$ & 0.544 & 1.309 & 128 \\ 
\hline 
PG 1630+377  & $2279-2980$ & $0.875-1.451$ & 0.577 & 1.396 & 118 \\ 
\hline 
PG 0117+213  & $2279-3009$ & $0.875-1.475$ & 0.600 & 1.460 & 160 \\ 
\hline 
HE 0515-4414 & $2278-3260$ & $0.874-1.682$ & 0.808 & 2.034 & 220 \\
\hline
HE 0141-3932 & $3061-3384$ & $1.518-1.784$ & 0.266 & 0.745 &  97 \\ 
\hline 
HE 2225-2258 & $3057-3478$ & $1.515-1.861$ & 0.346 & 0.979 & 130 \\ 
\hline 
\raisebox {-1ex}[-5ex]{HS 0747+4259}
& $2140-2970$ & $0.760-1.443$ & 0.683 & 1.615 & 110 \\ 
& $3115-3484$ & $1.562-1.866$ & 0.304 & 0.864 & 79  \\ 
\hline 
HE 0429-4901 & $3188-3538$ & $1.622-1.910$ & 0.288 & 0.830 &  88 \\ 
\hline
\hline
\hspace{1cm} $\Sigma$ & & & 5.176 & 12.897 & 1325 \\ 
\hline 
\end{tabular} 
\caption[] 
{\label{TabzX} Evaluated redshift ranges and absorption distances 
of the quasar LOS and numbers of Ly $\alpha$ lines $n$ detected
in the indicated spectral regions} 
\end{center}
\end{table*}

\subsection{Definition of redshift ranges} 

The sensitivity variations with wavelength of the individual detectors 
slightly reduce the redshift regions in which the {\Lya} forest 
can be examined. For HS~0747+4259, we defined a minimum signal to noise 
per pixel, averaged over each spectral order, for inclusion in the analysis. 
This is because the sensitivity of the HIRES chip decreases strongly
with decreasing wavelength. Choosing $\langle S/N\rangle > 2$, 
we ignored the first three orders for HS~0747+4259. Analogously,
the first two orders for HE~0429-4901 were left out. 
The spectrum of PG~1634+706
contains a Lyman limit system, making any detection of {\Lya} lines 
below $\lambda < 1865$ \AA\, impossible. We will discuss below how we separate 
{\Lya} from other Lyman lines.

We also excluded redshift regions within $\Delta v \sim 2\,500-5\,000$~{\kms} 
of the emission redshifts where the ionization of hydrogen is enhanced 
by the proximity effect. We calculated the maximum redshifts, 
given in Table \ref{TabzX}, from the absolute magnitudes of the QSOs.

For HS~0747+4259, there is a gap of approximately 145~\AA\, between the UV 
and the optical spectra, while for HE~0515-4414 there is an overlap of roughly 
50~\AA. In this overlap, we used only lines which can be identified 
in {\it both} spectra (nine altogether) and we chose the line parameters 
from UVES because they are more accurate. Six lines are only found in 
one spectrum (three in the STIS and three in the UVES spectrum, respectively),
all with column densities $\leq 10^{13.20}$ cm$^{-2}$. We suppose that
these are predominantly artefacts.

Absorbers with constant comoving number density and constant proper sizes
will have a changing density per unit redshift. We define the absorption
distance $\Delta X$ so that non-evolving absorbers have a constant density
per unit $X$ \citep{Tytler1982}. Assuming the $\Lambda $CDM model,

\begin{equation} 
\label{Boks} 
\Delta X = \frac{1+z}{\sqrt{\Omega_{{\rm M}}(1+z)+\frac{\Omega_\Lambda} 
{(1+z)^{2}}}}\,\, \Delta z 
\end{equation} 
\citep{Misawa2002b}, where $\Omega_{\rm M}$ and $\Omega_{\Lambda}$ 
are the density parameters for matter and for dark energy, respectively.

Given a set of QSO spectra, in general there are multiple LOS which sample
a specific redshift range. We take this into account by converting the 
$z$ into $X$ values and summing the $\Delta X$ values to obtain the total
absorption distance.

Table \ref{TabzX} shows the wavelength and redshift regions (columns two and
three) that we investigated for all quasars after removal of the proximity
zones and low $S/N$ orders. The fourth and fifth columns give the computed
redshift path $\Delta z$ and the absorption distance $\Delta X$. The total
absorption distance for all nine examined quasar LOS is $\Delta X=12.897$.
In the sixth column we give the number of detected {\Lya} absorption lines, 
1325 altogether.

\begin{figure*}
\begin{center}
\includegraphics[angle=90,scale=0.5]{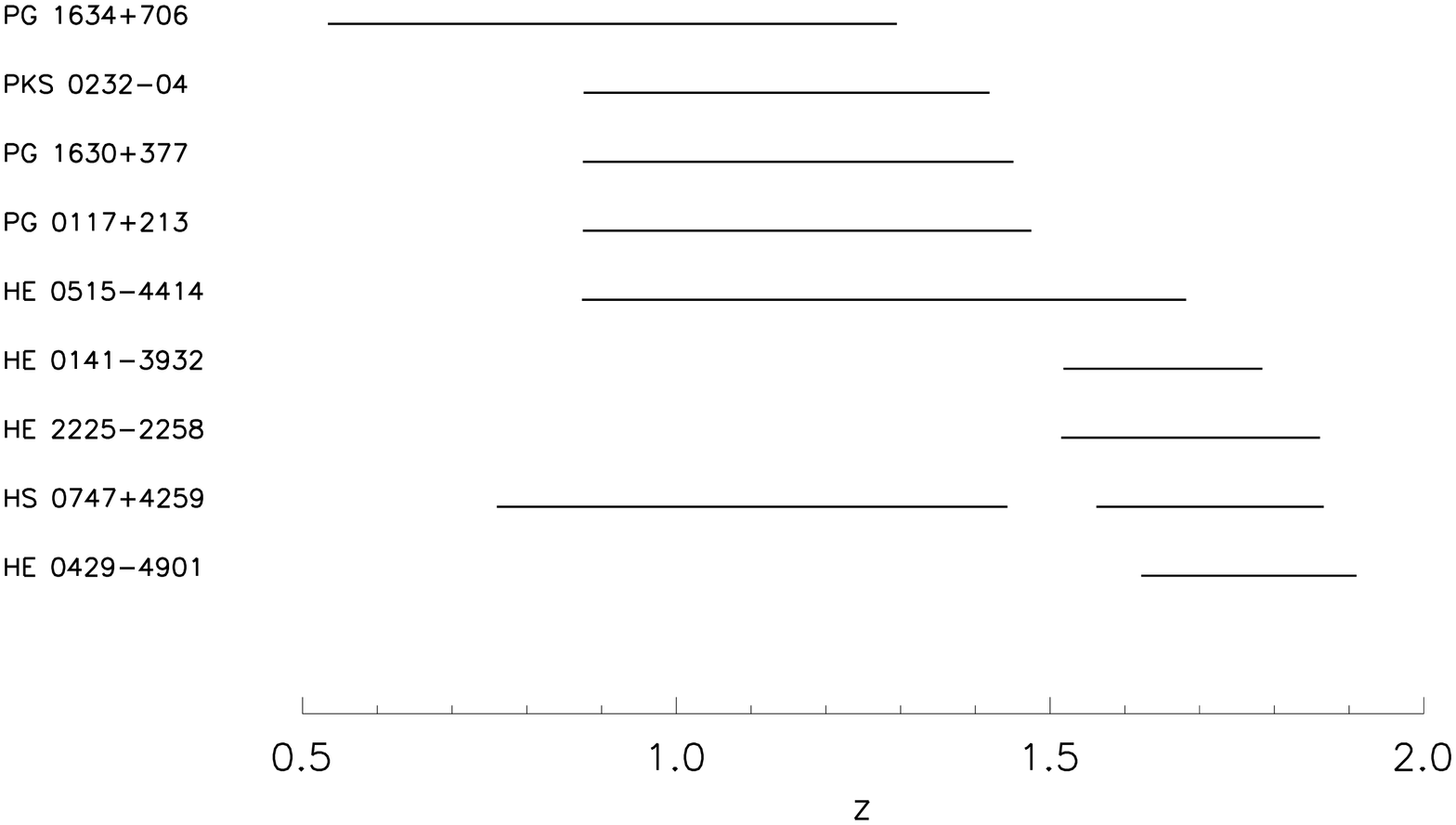}
\end{center}
\caption{}
{\label{LOS} Visualization of the analyzed redshift ranges of our quasars}
\end{figure*}

\begin{figure}[h]
\begin{center}
\includegraphics[angle=90,scale=0.35]{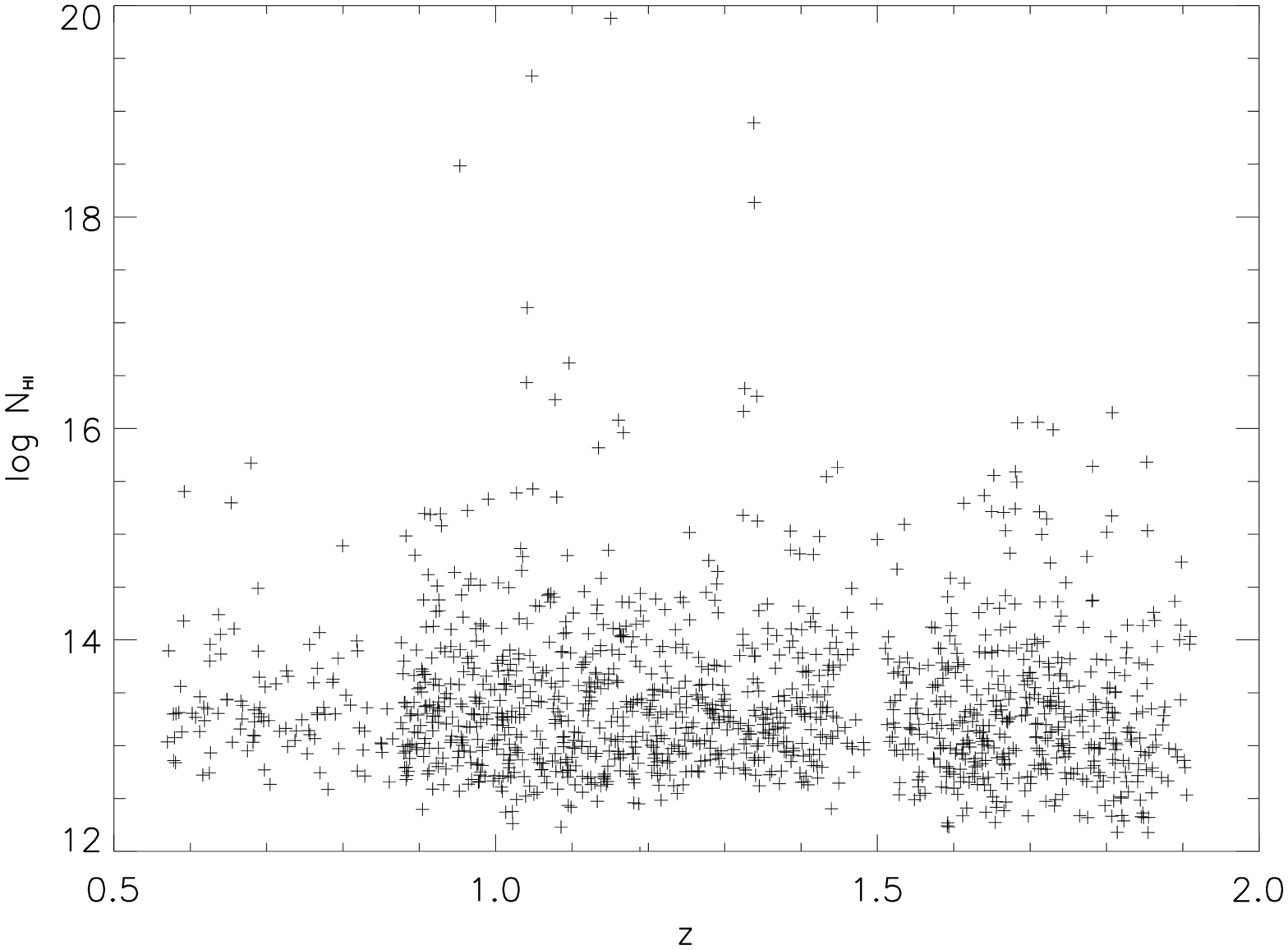}
\end{center}
\caption{}
{\label{Nz} Distribution of all 1325 {\Lya} lines as a function 
of redshift and column density}
\end{figure}

In Fig. \ref{LOS} we show the redshift intervals of the LOS, 
while in Fig. \ref{Nz} we show the distribution of all 1325 {\Lya} lines 
as a function of redshift and column density, giving qualitative information 
on the contents of the sample. It can be clearly seen 
that the redshift interval analysed in this paper is well sampled.

\section{Voigt profile fits}

The analysis of absorption lines detected in quasar spectra is usually
performed in three steps. First the lines have to be identified 
as certain atomic transitions that have been redshifted 
to the observed wavelengths. Then they are modelled. Finally 
they are accepted or rejected for the line sample.

For the line identification we used the empirical fact that the H{\sc i}
Doppler parameter $b$ is rarely smaller than 15 {\kms} and never 
$\leq 10$ {\kms}, while for metal lines almost always $b\leq 10$ {\kms}.
In most cases, it can a priori be discriminated between H{\sc i} 
and metal lines by inspecting the line width.

The identification strategy was as follows. After rejecting the interstellar
lines, the strongest {\Lya} absorbers (i.e., the lines with
the highest column densities) were picked out. Having identified a {\Lya}
line at the wavelength $\lambda_{\,{\rm Ly}\alpha}$ with a H{\sc i}
column density {\logn} $\gtrsim 14.00$ (the exact lower bound depending 
on the $S/N$), the associated Ly $\beta$ can in principle be found 
at the same redshift. For higher column density systems, further transitions 
of the Lyman series (with monotonically
decreasing line strength) can be detected.

Using the redshifts of the strongest H{\sc i} systems, we could search
for metal lines. We found many additional metal lines at the redshifts of
the common C{\sc iv} 1548/1550 doublets.

The absorption line list derived in this way still contains a
few unidentified lines, noise dips interpreted as absorption lines by
mistake, unidentified blends, and wrong identifications. However,
at these low redshifts, and especially in the high-resolution UVES spectra 
with their high $S/N$ we guess that such errors are less than 5\% 
of the total list. Besides, the unidentified or wrongly identified lines are
typically those with very low column densities and/or low Doppler
parameters, which are either rejected by applying the significance level 
(see below) or do not belong in our {\Lya} line sample.
On the other hand, the blends are a problem that is easily underestimated.

We fit the {\Lya} lines with two different programs:
the lines found in the HST/STIS and VLT/UVES spectra of HE~0515-4414 were
fit with the MIDAS software package FITLYMAN \citep{Fontana}, which needs
a normalized spectrum as input. In order to obtain this, the effective
background continuum of the quasar was defined with MIDAS by specifying
points in line-free regions and fitting these by a polynomial. An
improvement to FITLYMAN is the program CANDALF developed by R.~Baade. In
CANDALF, the continuum is determined {\it simultaneously} with the line
fitting procedure. With this parallel approach, the continuum can also be
defined reliably in spectral regions where it is hidden by a high line
density. CANDALF was used for the analysis of all quasar spectra except for
HE~0515-4414 whose spectra were already fit before the development of
CANDALF.

Extensive investigations of complex absorption line ensembles, several with
partly blended atomic transitions and lines of multiplets fit
simultaneously, were carried out with both programs to check their
consistency. The comparison of the ionic parameters determined with FITLYMAN
and with CANDALF showed that they were consistent within $1\sigma$ 
in almost all cases. CANDALF converges more easily compared with
FITLYMAN in regions with high line densities, finds the global fit minimum
with larger probability, and its fit errors are a little bit smaller;
however the choice of the fitting program does not have any significant
influence on the results of the analysis of the whole {\Lya}
sample (Section \ref{discussion}).

FITLYMAN and CANDALF both assume Voigt profiles convolved with the
instrumental profile. The programs adjust three independent parameters per
line to minimize the $\chi^{2}$: $z$, $N_{\rm \ion{H}{i}}$
and $b$ which comprises the thermal and the turbulent broadening of the lines. 
The general fit strategy, especially for blends,
was to start with a single line and to add further components as long as
the $\chi^{2}$ decreased significantly. Different atomic transitions of the
same ion were fit with identical values for $N_{\rm \ion{H}{i}}$, $b$ and $z$. 
This is particularly important for the saturated {\Lya} lines 
({\logn} $\gtrsim 13.50-14.00$) lying in the flat part of the curve of growth 
where log~$N_{\rm \ion{H}{i}}$ is very inaccurate. In addition the number of
subcomponents of an absorber is harder to determine at higher column
densities. However, the fragmentation of an absorber into several weak
absorbers can be seen in Ly $\beta$ and other Lyman series lines because
these lines are usually not saturated and give more accurate parameters.

Since the lower column density lines can hardly be distinguished from the
photon noise, it is usual practice to define the significance level

\begin{equation} 
SL = \frac{W}{\sigma_{W}}
\end{equation} 
where $W$ is the observed equivalent width of the line and $\sigma_{W}$ 
is the $1\sigma$ error of $W$ \citep{Young1979}. Usually the approximation 
$\sigma_{W}\sim \frac{FWHM}{\langle S/N\rangle }$ with $\langle S/N\rangle$ 
being the average signal to noise in the region of the considered 
absorption line is used \citep{Tytler1987, Caulet}. This describes the error 
caused by the determination of the continuum. However it neglects 
the fit error \citep{Sembach}. Here we performed an extended estimation 
of $SL$ that includes the fit error:

\begin{equation}
\sigma_{W} \sim \sqrt{\left(\frac{\lambda}{R\, \langle S/N\rangle}\right)^{2} 
+ \sigma_{\rm Fit}^{2}}.
\end{equation}
Because of this extended error estimation, we chose a comparatively 
low selection threshold $SL>1$. With this selection criterium
about 5\% of all presumed {\Lya} lines were rejected from the sample.
However, we did not reject {\Lya} lines that are at redshifts 
where metal lines can be seen. 

The full list of all 1325 detected and modelled {\Lya} lines 
with their fit parameters and errors are available in electronic form 
at the Centre de Donn\'{e}es astronomiques de Strasbourg (CDS) data base 
via anonymous ftp to cdsweb.u-strasbg.fr.

\section{Individual LOS}

Before we analyze the {\Lya} line sample from all nine quasar LOS 
in Section \ref{discussion}, we briefly mention some interesting
characteristics in the individual LOS.

\subsection{HE~0515-4414} 

In the LOS to the quasar HE~0515-4414 a damped {\Lya} (DLA) system can be seen
at $z \sim 1.15$. Because of this absorber and the brightness of the quasar 
($m_{B}=15.0$ mag), it was already a topic of several studies with
different goals by the Hamburg group
\citep{Varga2000, Reimers2001, Reimers2003, Janknecht2002, 
Quast2002, Quast2004, Quast2006, Levshakov2003}.
  
\subsection{HS~0747+4259}

The metal line spectrum of HS~0747+4259 has been studied in a separate paper
\citep{Reimers2006} which concentrated on the O{\sc vi}
absorbing systems in this LOS.

\subsection{HE~0141-3932} 

HE~0141-3932 is an unusual quasar in several respects: its spectrum contains
--- atypical for quasars --- an extremely weak {\Lya} emission line. 
In addition the emission lines of different ions give rather different 
redshifts, depending on the degree of ionization (e.g., Tytler \& Fan, 1992). 
Furthermore, the quasar has several associated absorption systems presumably 
built by gas ejected from the QSO. The chemical composition of this gas 
is rather atypical. These special characteristics of HE~0141-3932 are discussed
in detail in \cite{Reimers2005} and in \cite{Levshakov2005}.


\section{Analysis and Discussion}
\label{discussion}

\subsection{Column density distribution}
\label{Ndis}

The differential distribution function of the hydrogen column densities 
$f(N_{\rm \ion{H}{i}})$ is usually defined as the number $n$ 
of {\Lya} absorption lines per column density interval 
$\Delta N_{\rm \ion{H}{i}}$ and per absorption distance 
$\Delta X$ ($\Delta X$ from (\ref{Boks}); Tytler, 1987). 
Since our data contains more than one LOS,
we need the sum of the individual $\Delta X_{i}$ over all LOS:

\begin{equation} 
f(N_{\rm \ion{H}{i}}) = \frac {n}{\Delta N_{\rm \ion{H}{i}} \,\,
\Sigma_{i=1}^{j}\Delta X_{i}}.  
\end{equation} 

\begin{figure} 
\includegraphics[angle=90,scale=0.35]{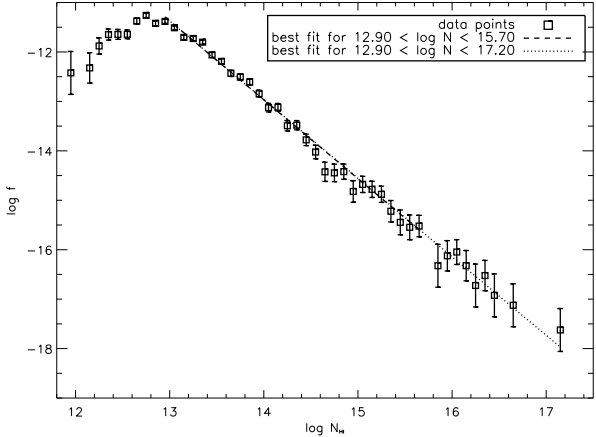} 
\caption[] 
{\label{cddf_12-17.5_0.1_sig_bw} Distribution of the {\Lya} {\Ns} in intervals 
$\Delta\,{\rm log}~N_{\rm \ion{H}{i}}=0.1$, statistical error 
of log~$f$ and fits to the distribution for different fit regions. 
For the calculation of the fit parameters, the data points were weighted 
with their error.}  
\end{figure}

In Fig. \ref{cddf_12-17.5_0.1_sig_bw} the distribution function for the
lines of all nine LOS is plotted against the column density. Note that the
error bars on log~$f$ increase with increasing {\N}
because of the decreasing number of lines. Fig. \ref{cddf_12-17.5_0.1_sig_bw}
shows that the data points at the highest {\Ns} have a low statistical 
significance. The fits plotted in this diagram were derived by weighting 
the data points with $1/\sigma_{{\rm log}~f}^{2}$. The absorption lines 
were binned in intervals $\Delta\, {\rm log}\, N_{\rm \ion{H}{i}}=0.1$. 
As expected, the linearity of log $f$ breaks off at low column densities. 
This presumably reflects a selection effect: the weak lines are hidden 
by the the noise of the spectra.

In general, the column density distribution function can be well approximated
by the power law 

\begin{equation}
\label{cddf}
f(N_{\rm \ion{H}{i}}) = A \cdot N_{\rm \ion{H}{i}}^{\,-\beta},
\end{equation}
where $\beta$ is the (negative) slope and log $A$ is the intercept 
of the logarithmic representation of $f(N_{\rm \ion{H}{i}})$.

We set a completeness limit \mbox{log~$N_{\rm \ion{H}{i}}=12.90$} for the
analyzed line ensemble, while we chose \mbox{log~$N_{\rm \ion{H}{i}}=15.70$}
for the upper limit (see Fig.~\ref{cddf_12-17.5_0.1_sig_bw}). Using these
boundaries and fitting the model (\ref{cddf}) to the column density
distribution, we derived \mbox{log $A=9.4\pm0.3$} and $\beta =1.60\pm 0.03$.
This result is consistent with other studies of the column density
distribution at lower log~$N_{\rm \ion{H}{i}}$ at comparable $z$: 
\cite{Dobrzycki} found \mbox{$\beta \leq 1.6 - 1.7$} 
performing a curve of growth analysis. \cite{Hu} determined $\beta =1.46$ 
for log~$N_{\rm \ion{H}{i}}=12.30-14.50$, and \cite{Kim2001} 
derived an exponent $\beta =1.70-1.74$ with varying column density regions.

Occasionally it has been claimed that at least two power laws are necessary
to accurately describe the {\N} distribution over the observed range 
of {\N} \citep{Carswell1987, Giallongo, Meiksin, Petitjean, Penton2004}. 
If we extend our fit region to higher {\Ns} 
\mbox{(log~$N_{\rm \ion{H}{i}}=12.90-17.20$)}, 
the distribution does not flatten ($\beta =1.59\pm 0.02$). 
However, our sample is too small at large column densities 
to see minor changes in this $N_{\rm \ion{H}{i}}$ region. 

Table \ref{beta} summarizes the results for the fit parameters log~$A$ and 
$\beta$ for different column density regions and interval widths.

\begin{table}
\begin{center}
\begin{tabular}{|c|c|r|c|}\hline
log $N_{\rm \ion{H}{i}}$ & $\Delta {\rm log} N_{\rm \ion{H}{i}}$ 
& log $A$\hspace{0.4cm} & $\beta$ \\
\hline
12.90--15.70 & 0.1 & $ 9.4\pm0.3$ & $1.60\pm0.03$ \\
12.90--17.20 & 0.1 & $ 9.3\pm0.3$ & $1.59\pm0.02$ \\
13.00--15.50 & 0.5 & $10.0\pm0.4$ & $1.64\pm0.03$ \\
13.00--16.50 & 0.5 & $ 9.9\pm0.3$ & $1.63\pm0.02$ \\
\hline
\end{tabular}
\end{center}
\caption[]
{\label{beta} Fit parameters of the column density distribution for different 
boundary conditions. Given are (from left to right): 
the log~$N_{\rm \ion{H}{i}}$ range of the data points considered in the fit; 
the interval width; the fit parameters log $A$ and $\beta$ 
with their 1$\sigma$ errors.}
\end{table}

\begin{figure} 
\includegraphics[angle=90,scale=0.35]{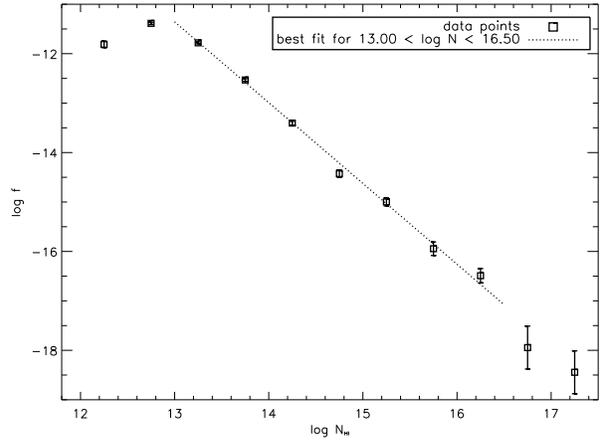} 
\caption[] 
{\label{cddf_12-17.5_0.5_sig_bw} Distribution of the {\Lya} {\Ns} in intervals 
$\Delta\, {\rm log} \,N_{\rm \ion{H}{i}}=0.5$, 
statistical error of log~$f$ and fit to the distribution. 
The data points were weighted with the error for the fit.}  
\end{figure}

Of course the choice of the size of the intervals 
$\Delta\,{\rm log} \,N_{\rm \ion{H}{i}}$ in which the lines are collected 
is somewhat arbitrary. Fig.~\ref{cddf_12-17.5_0.5_sig_bw} is based 
on a larger step size $\Delta\,{\rm log}\, N_{\rm \ion{H}{i}}=0.5$. 
For \mbox{log~$N_{\rm \ion{H}{i}}=13.00-16.50$}, the distribution 
can be well approximated by $\beta=1.63\pm0.02$ (again weighted 
with $\sigma^{2}_{{\rm log} f}$).
This is within 1$\sigma$ consistent with the slope for the interval width 0.1 
in the comparable range log~$N_{\rm \ion{H}{i}}=12.90-17.20$ 
(only two lines lie in the interval log~$N_{\rm \ion{H}{i}}=16.50-17.20$), 
which makes the point that the choice of the interval width does not
influence the result.

Frequently a dependence of the index $\beta$ on $z$ has been postulated by 
observers \citep{Kim1997, Kim2001, Kim2002, Dave2001, Heap, Misawa2002}.
\cite{Dave2001} found $\beta=2.04\pm0.23$ and \cite{Heap} derived
$\beta=2.02\pm0.21$ for $z \sim 0$ each, while \cite{Kim2001} 
calculated $\beta=1.72\pm0.16$ for $\langle z\rangle=1.6$ and 
$\beta=1.38\pm0.08$ for $\langle z\rangle=2.1$, \cite{Telfer2002} got
$\beta=1.41\pm0.05$ for $\langle z\rangle \sim 2.3$ and 
\cite{Hu} and \cite{Kim1997} found $\beta=1.46$ and $\beta=1.4$, respectively,
both for $\langle z\rangle \sim 3$ (all values calculated for the low 
{\N} region log~$N_{\rm \ion{H}{i}} \leq 14$ where the number of lines 
make a reliable judgement possible). Simulations of \cite{Theuns1998}, 
e.g., suggested a clear relationship $\beta(z)$, too.  
These results are contradicted by the comparatively low gradients 
of Penton et al. (2000, 2004) for the local universe 
($\beta=1.72\pm0.06$ and $\beta=1.65\pm0.07$, respectively).  
However, the difference might be caused by the analysis method 
since Penton et al. (2000, 2004) assumed a constant {\B} for the line fits.

\begin{figure} 
\includegraphics[angle=90,scale=0.35]{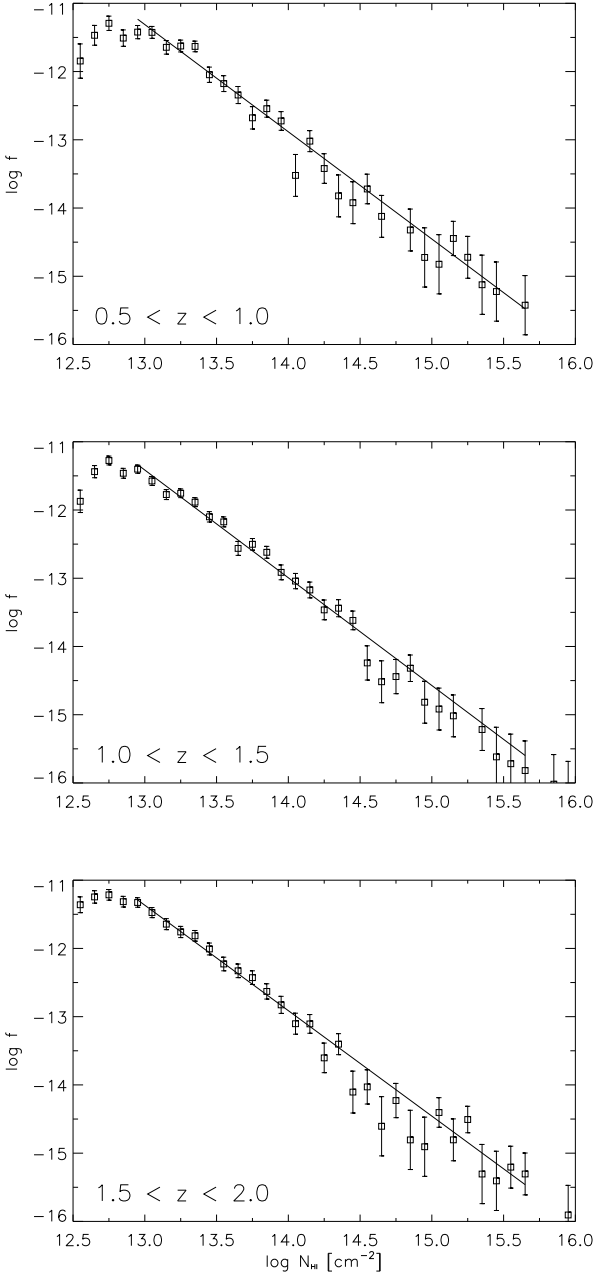} 
\caption[] 
{\label{cddf_ev_sig} Distributions of the {\Lya} {\Ns} in intervals 
\mbox{$\Delta\,{\rm log}\, N_{\rm \ion{H}{i}}=0.1$}, statistical errors 
of log~$f$ and fits to the distributions for different {\z} intervals. 
Only the data points within the range log~$N_{\rm \ion{H}{i}}=12.90-15.70$ 
are considered for the fits; they are weighted with the error of log~$f$.}  
\end{figure} 

In order to search for a possible dependence $\beta(z)$, we examined 
the {\cddf} separately for three $z$
intervals of the same width ($0.5-1.0$, $1.0-1.5$, $1.5-2.0$). For this
purpose, the LOS to the quasars had to be divided according to the chosen $z$
intervals, and the absorption distance $\Delta X$ had to be computed for the
individual intervals following (\ref{Boks}). We selected 
$\Delta\, {\rm log}\, N_{\rm \ion{H}{i}}=0.1$ for the interval width. 
The distributions for all three {\z} regions and the fits to them are presented
in Fig. \ref{cddf_ev_sig}. In each case the fits take into account the
points in the interval log~$N_{\rm \ion{H}{i}}=12.90-15.70$. Obviously,
the fit parameters given in Table \ref{beta_ev} are independent of the
redshift range. The increase of $\beta $ from $z=1.5-2.0$\, 
($\beta=1.55\pm0.04$) to $z=1.0-1.5$\, ($\beta=1.58\pm0.04$)
represents at best a marginal trend which is not statistically significant.

\begin{table} 
\begin{center} 
\begin{tabular}{|c|c|r|c|}
\hline 
$z$ & log~$N_{\rm \ion{H}{i}}$ & log~$A$ \hspace{0.4cm} & $\beta$ 
\\ 
\hline 
$0.5-1.0$ & $12.90-15.70$ & $ 9.1\pm0.7$ & $1.57\pm0.05$ \\ 
$1.0-1.5$ & $12.90-15.70$ & $ 9.1\pm0.5$ & $1.58\pm0.04$ \\ 
$1.5-2.0$ & $12.90-15.70$ & $ 8.7\pm0.6$ & $1.55\pm0.04$ \\ 
\hline 
\end{tabular} 
\end{center} 
\caption[] 
{\label{beta_ev} Fit parameters of the column density distribution
for different redshift ranges. Indicated are (from left to right):  
the $z$ interval; the log~$N_{\rm \ion{H}{i}}$ range of the data points 
considered in the fit; the fit parameters log~$A$ and $\beta$ 
with their 1$\sigma$ errors.}  
\end{table}

\subsection{Distribution of Doppler parameters}

\begin{figure} 
\includegraphics[angle=90,scale=0.35]{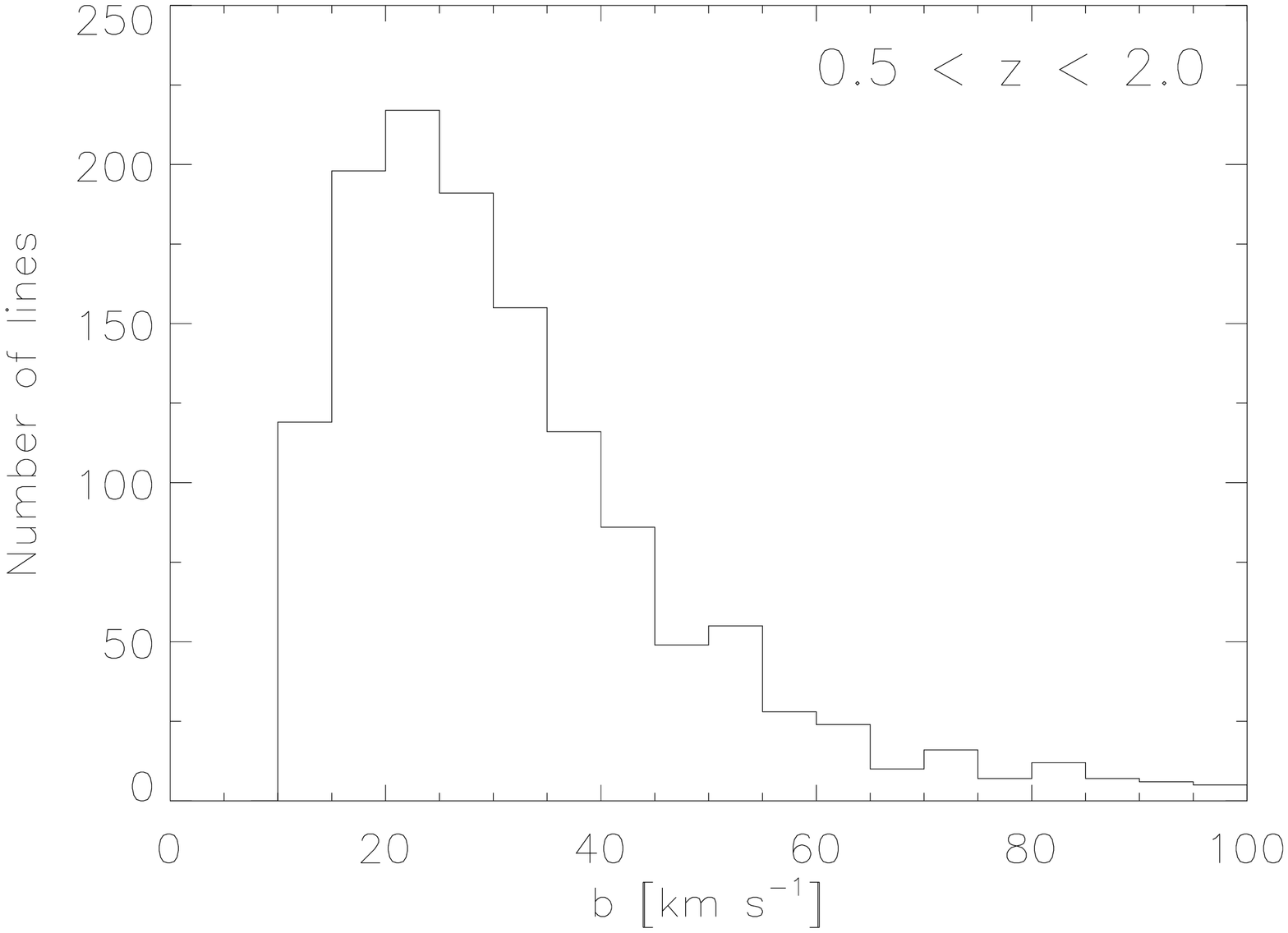} 
\caption[] 
{\label{bdis} Distribution of the {\Lya} Doppler parameters in intervals 
\mbox{$\Delta\, b=5$ {\kms}}.}
\end{figure}

To investigate the distribution of the Doppler parameters, we counted the
number of lines per Doppler parameter interval in interval widths 
$\Delta\, b=5$ {\kms}. Fig. \ref{bdis} shows how $b$ is distributed
across the complete line sample (without the 24 broadest lines with  
$b>100$ {\kms}). The distribution has a typical form: nearly Gaussian 
with a maximum at $b=(20-25)$~{\kms}, an average value 
$\overline{b}=(34\pm 22$) {\kms} lying above the maximum, and a long tail 
to higher Doppler widths. A large fraction of the lines with 
$b>100$ {\kms} might represent unresolved blends of several components, 
whereas the steep decline at small Doppler parameters is caused 
by the $SL$ limit (lines with small $b$ values have small equivalent widths 
and are more likely to be missed) and by our {\Lya} selection threshold 
$b>10$~{\kms}.

The average and median of $b$ fit the results from other studies in
comparable {\z} regions very well: \cite{Kim2001} found $b_{\rm median}=28$ 
{\kms} for $\langle z\rangle = 1.6$, while \cite{Kim2002} derived 
$\overline{b} = 32.6$ {\kms} for $\langle z\rangle = 2.2$.

Frequently a dependence $b(z)$ is claimed in the sense of a Doppler width
increasing with decreasing {\z}. In the high redshift range \cite{Lu} found 
$\overline{b} = (23 \pm 8)$ {\kms} for $\langle z\rangle =3.7$, \cite{Kim1997} 
derived $b_{\,\rm median}=26$ {\kms} for $\langle z\rangle = 3.35$
and $b_{\,\rm median}=30$ {\kms} for $\langle z\rangle = 2.31$, respectively,
and \cite{Hu} found $\overline{b} = (28 \pm 10)$~{\kms} for 
\mbox{$\langle z\rangle = 2.9$}. The average value of this study for 
$\langle z\rangle =1.31$ appears to support this trend. Going to $z=0$, 
the evolution of $b$ possibly continues: \cite{Penton2000} derived 
$\overline{b}=(38\pm16)$ {\kms} for the local universe.

\begin{figure} 
\includegraphics[angle=90,scale=0.35]{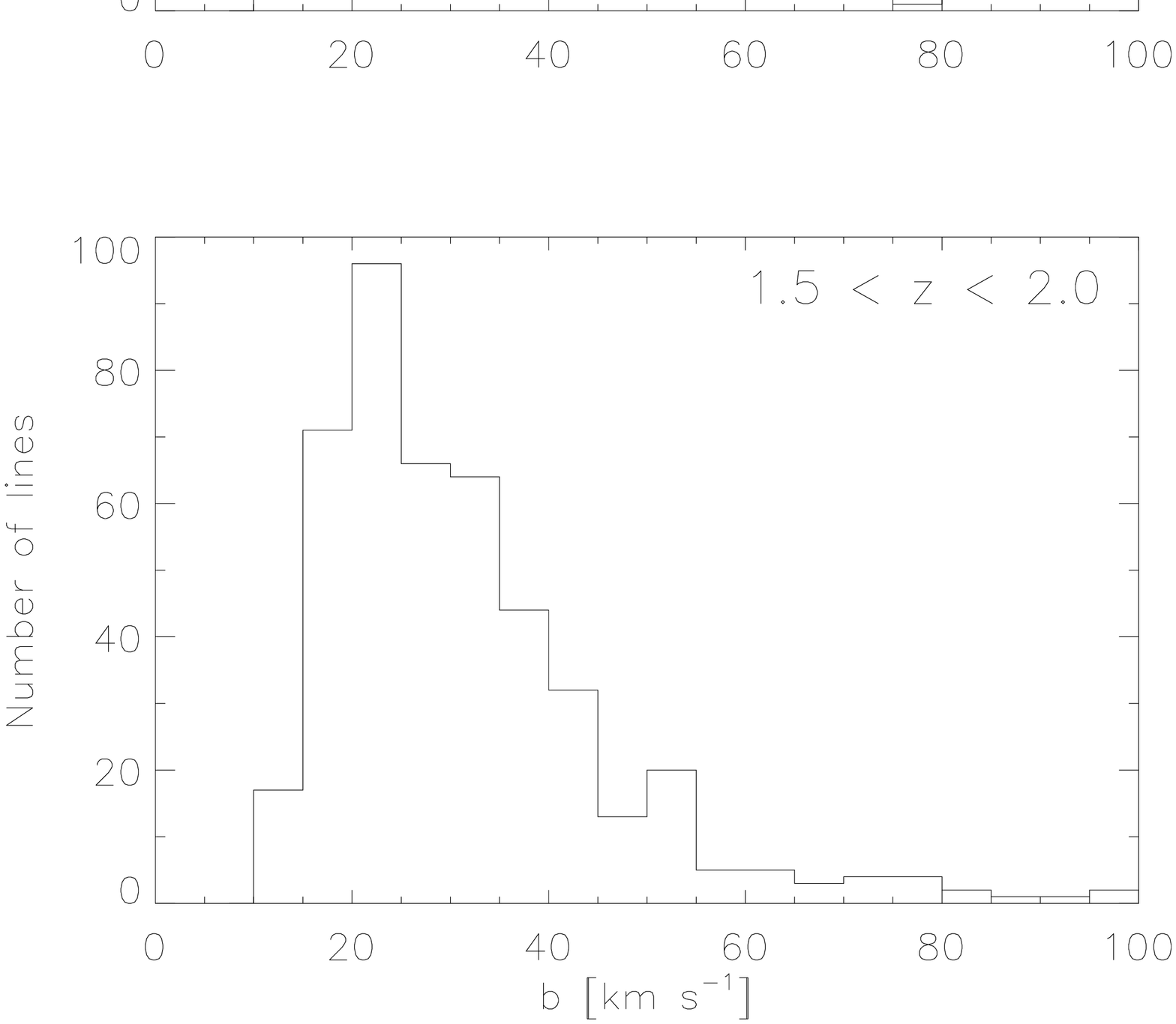} 
\caption[] 
{\label{bdisz} Distribution of the {\Lya} Doppler parameters in intervals 
\mbox{$\Delta\, b=5$ {\kms}}, varying {\z} ranges.}
\end{figure} 

\begin{table}
\begin{center}
\begin{tabular}{|c|r|c|c|c|}
\hline
$z$ & $n$ & $\overline{b}$ \,[{\kms}] & $b_{\,\rm median}$ [{\kms}] & 
$b_{\sigma}$\, [{\kms}] \\
\hline
$0.5-2.0$ & 1325 & $34\pm22$ & 28 & 22.7 \\ 
$0.5-1.0$ & 270 & $35\pm23$ & 29 & 18.9 \\ 
$1.0-1.5$ & 595 & $33\pm22$ & 28 & 23.5 \\ 
$1.5-2.0$ & 460 & $34\pm22$ & 28 & 23.5 \\ 
\hline
\end{tabular}
\end{center}
\caption{Parameters of different $b$ distributions. We give
the number of {\Lya} lines detected, the average value, the median of $b$, 
and on the right a parameter that represents the mode.}
\label{bmit+med}
\end{table}
 
Therefore we investigated whether an evolution of $b$ with $z$ can be
determined within the $z$ range examined here. In Fig.~\ref{bdisz} the $b$
distribution is shown for three different $z$ ranges of the same width,
while Table \ref{bmit+med} gives parameters of the distributions. 
The highest average value \mbox{$\overline{b}=(35\pm23)$~{\kms}} 
for the interval $z=0.5-1.0$ lies only marginally above those in the
higher {\z} regions. We see no evolution in the $b$ mean and
median over redshifts 0.5 to 2. In general, the question of whether 
the Doppler width changes with $z$ is not resolved: \cite{Kim2002} also did not
see any evolution of $\overline{b}$ in the interval $\langle z\rangle
=3.3\rightarrow 2.1$ in their data, nor did \cite{Dave2001} ($\overline{b}
=25 $ {\kms} for $\langle z\rangle =0.2$), whose results do not fit 
the evolution pattern outlined above.

In addition to the mean and median, we also list a parameter 
that represents the mode, or most common $b$ value, because this tends to be
less sensitive to the $S/N$ and the details of line and continuum fitting.
We list values for the parameter $b_{\sigma} = 1.0574 \cdot b_{\rm peak}$
from \cite{Hui99}. \cite{Tytler2004} and \cite{Jena2005} found 
that the \cite{Hui99} fitting formula with a single parameter 
$b_{\sigma}=(23.6\pm1.5)$~{\kms}
gives excellent fits to the $b$ distribution at $1.5 < z < 2.4$. 
Here we obtain essentially the same value for our two higher redshift bins, 
1.0 -- 1.5 and 1.5 -- 2.0, and a smaller value at lower redshifts 0.5 -- 1.0. 
The $b$ value distribution is significantly wider than the fitting function 
for the two lower redshift bins, and hence these two $b_{\sigma}$ values 
are not well determined. Both these two lower redshift bins show 
many more lines with $b < 15$ {\kms} than we see at $z > 1.5$. We know, 
for example from \cite{Kirkman2005}, that as the $S/N$ of a spectrum drops,
as it does here dramatically at low $z$, we will measure an excess 
of both low and high $b$ values. Hence, all the changes that we see may
arise from the changing $S/N$. We do not claim to detect any change 
in the {\it intrinsic} $b$ distribution with $z$.

\subsection{Clustering of {\Lya} absorbers}
\label{clust}

The clustering properties of the baryonic matter can be investigated by
computing the correlation between the {\Lya} absorbers along
the LOS. The degree of clustering of the \ion{H}{i} absorbers can be
expressed by the two-point velocity correlation function

\begin{equation} 
\xi\,(\Delta v) = \frac{n_{\rm obs} \,(\Delta v)}{n_{\rm sim}\,(\Delta v)} -1 
\end{equation} 
\citep{Sargent} where $n_{\rm obs}$, the number of observed {\Lya} line pairs 
in a given velocity separation bin $\Delta v$, is compared with the number 
of expected line pairs $n_{\rm sim}$ derived in the same velocity difference 
bin in a spectrum with randomly placed lines. For two absorbers 
at the {\z}s $z_{1}$ and $z_{2}$, the velocity interval at mean redshift 
in the rest frame is given by

\begin{equation} 
\Delta v = \frac{c\, (z_{2}-z_{1})}{1+\frac{z_{1}+z_{2}}{2}}.
\end{equation} 

We determined $n_{\rm sim}$ from Monte Carlo simulations by
distributing our line sample --- in accordance with the derived number
density evolution (see Subsection \ref{nde}) --- randomly over the spectrum,
counting the line pairs for various velocity splittings, repeating this
procedure 1000 times, and computing the average.

When the clustering properties in multiple LOS are investigated, each quasar
spectrum must first be analyzed separately. Subsequently, the individual
numbers of pairs $n_{\rm obs}$ and $n_{\rm sim}$ of the $j$ analyzed LOS 
can be added in order to build a total $\xi$ based on improved statistics:

\begin{equation} 
\label{xitot}
\xi\,(\Delta v) = \frac{\sum_{i=1}^{j} n_{{\rm obs},i}\, (\Delta v)}
{\sum_{i=1}^{j} n_{{\rm sim},i}\, (\Delta v)} -1.  
\end{equation}

\begin{figure} 
\includegraphics[angle=90,scale=0.35]{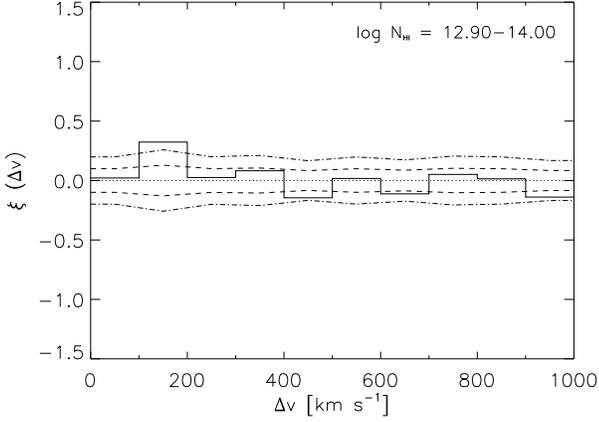} 
\caption[] 
{\label{tpcf_wl_1000_all} Two-point velocity correlation function 
for the weaker absorbers, up to 1\,000 {\kms}, in 100 {\kms} bins. 
The correlation function is shown by the solid line. For orientation 
$\xi=0$ is marked with a dotted line. Dashed and dot-dashed lines 
represent the $1 \sigma$ and $2 \sigma$ Poisson errors, related to $\xi=0$, 
respectively.}  
\end{figure} 

\begin{figure} 
\includegraphics[angle=90,scale=0.35]{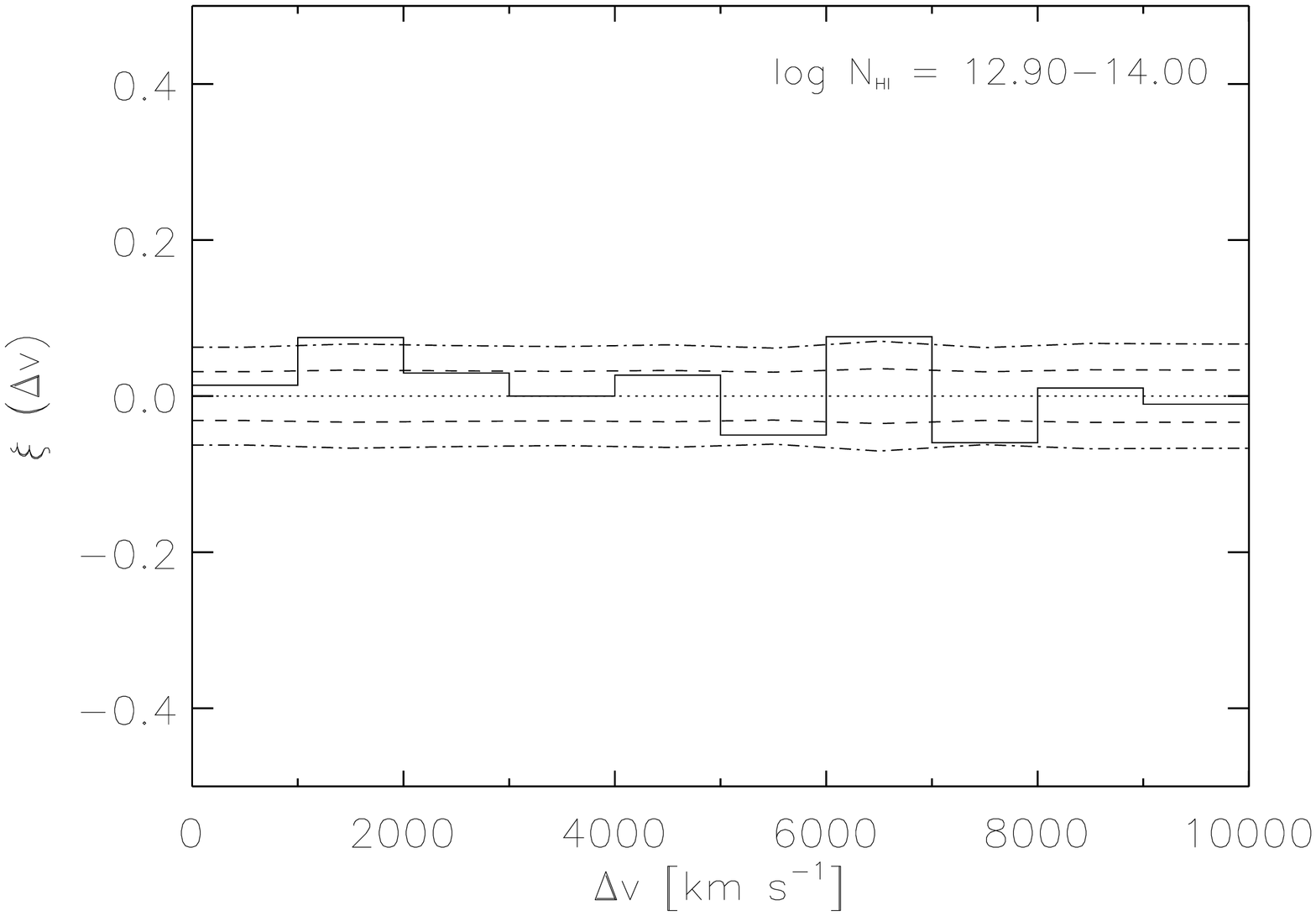} 
\caption[] 
{\label{tpcf_wl_10000_all} Two-point velocity correlation function 
for the weaker absorbers, up to 10\,000 {\kms}, in 1\,000 {\kms} bins. 
Line symbols are the same as in Fig. \ref{tpcf_wl_1000_all}.}  
\end{figure} 

\begin{figure}[h]
\includegraphics[angle=90,scale=0.35]{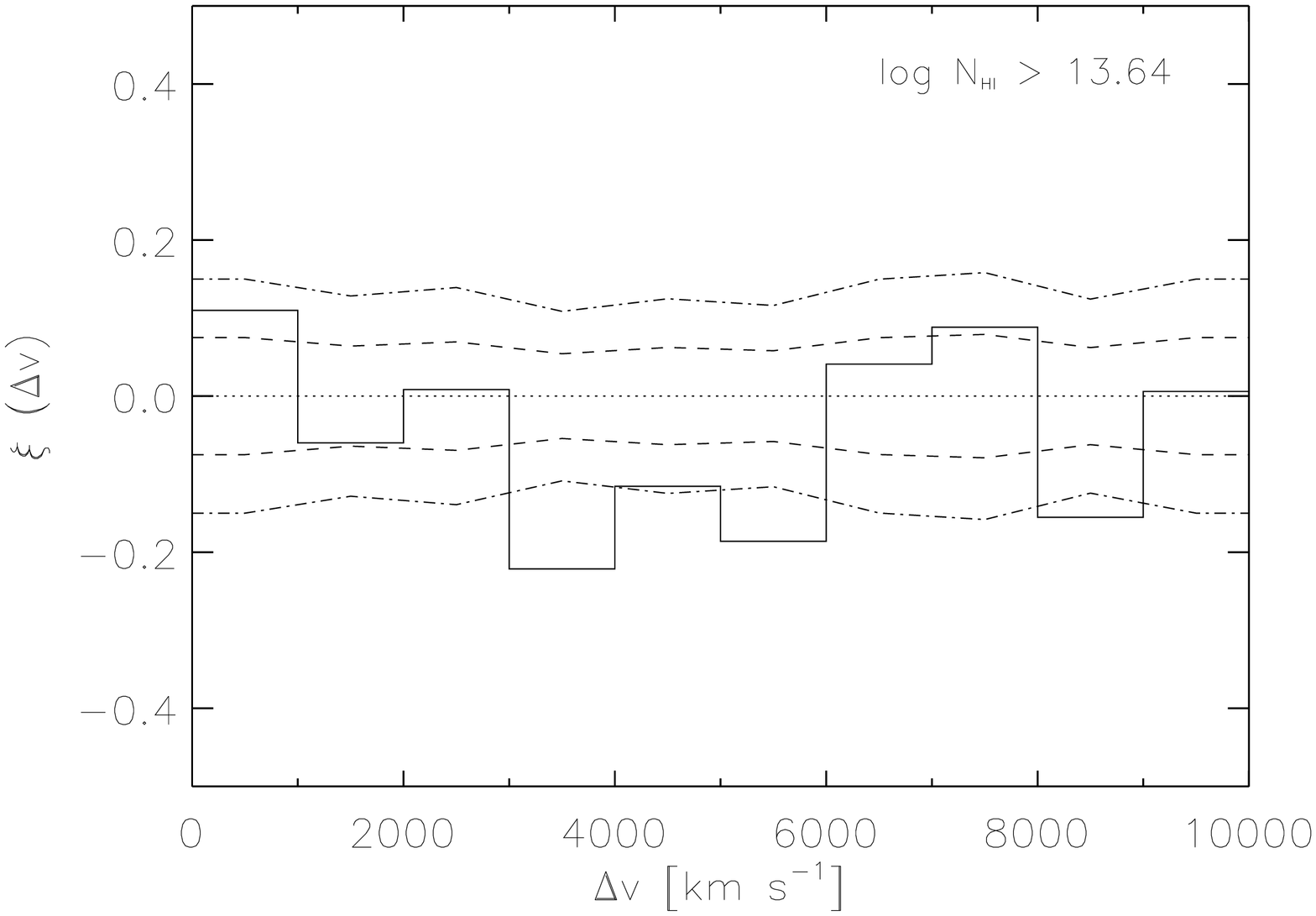} 
\caption[] 
{\label{tpcf_sl_10000_all} Two-point velocity correlation function 
for the stronger absorbers, up to 10\,000 {\kms}, in 1\,000 {\kms} bins.
Line symbols are the same as in Fig. \ref{tpcf_wl_1000_all}.}  
\end{figure} 

\begin{figure}[h]
\includegraphics[angle=90,scale=0.35]{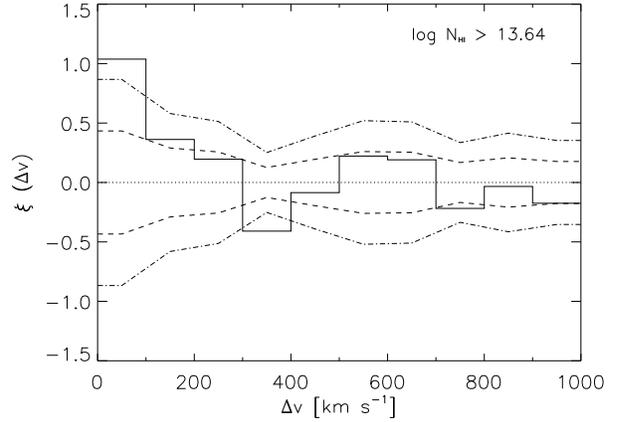}
\caption[] 
{\label{tpcf_sl_1000_all} Two-point velocity correlation function 
for the stronger absorbers, up to 1\,000 {\kms}, in 100 {\kms} bins.
Line symbols are the same as in Fig. \ref{tpcf_wl_1000_all}.}  
\end{figure} 

For varying intervals $\Delta z$ and $\Delta\, {\rm log}\, N_{\rm \ion{H}{i}}$,
we counted $n_{\rm obs}$ and simulated $n_{\rm sim}$ for every LOS. 
To investigate the clustering on small scales, we chose 
$\Delta v_{\rm max}=1\,000$ {\kms} using a binning of 100 {\kms}, 
and we defined $\Delta v_{\rm max}=10\,000$ {\kms} (in 1\,000~{\kms} bins) 
for a study of the long-scale clustering properties. We calculated $\xi$
according to (\ref{xitot}) with $j=10$ (because the LOS to HS~0747+4259 was
formally treated as two LOS: one for $z<1.44$ and one for $z>1.56$; compare
with Table \ref{TabzX}). We estimated the error on the clustering
signal with error propagation, assuming a Poisson error for $n_{\rm sim}$ 
and setting $\sigma _{n_{\rm obs}}=0$. We think that this is an
acceptable approximation because $\sigma _{n_{\rm obs}}$ follows
directly from $\sigma _{\lambda}$ or $\sigma_{z}$, respectively, and the
line positions are very well known compared with the values of the
investigated line separations $\Delta z$. Furthermore, $n_{\rm obs}$ and 
$n_{\rm sim}$ have similar values in most cases.

Since Cristiani et al. (1995) found that clustering is stronger for higher
column density absorbers, we also looked for this trend. We took the
conventional definition log~$N_{\mathrm{\ion{H}{i}}}>13.64$ for the strong
lines, and we chose log~$N_{\mathrm{\ion{H}{i}}}=12.90-14.00$ for the low {\N}
absorbers in accordance with the completeness limit of our {\cddf} 
(see Subsection \ref{Ndis}).

In Figs. \ref{tpcf_wl_1000_all} and \ref{tpcf_wl_10000_all} we show 
$\xi (\Delta v)$ for the low {\N} absorbers on scales up to 1\,000 {\kms} 
and 10\,000 {\kms}, respectively. We detect a marginal signal 
above the $2 \sigma$ level
only for $\overline{\Delta v}=150$~{\kms}, 1\,500~{\kms} and 6\,500~{\kms}.
Apart from that, $\xi(\Delta v)$ is always below the $1 \sigma$ level.
The excess at \mbox{$\overline{\Delta v}=150$~{\kms}} 
($\xi=0.32$, $1 \sigma=0.13$) suggests a clustering of the weak lines 
over short distances. The signal at $\overline{\Delta v}=50$~{\kms} 
($\xi=0.02$, $1\sigma=0.10$) might be too weak because some weaker lines 
remain unidentified due to blends: resolving these blends into superpositions
of several lines would result in far more line pairs at low $\Delta v$.

As expected, we also found slight clustering signals 
for the high column density absorbers, however only over short distances: 
$\xi = 1.04 > 2 \sigma$ for \mbox{$\overline{\Delta v}=50$~{\kms}} and 
$\xi = 0.36 > 1 \sigma$ for $\overline{\Delta v}=150$~{\kms} 
(Fig. \ref{tpcf_sl_1000_all}). At higher distances (up to $1\,000$~{\kms}),
the strong absorbers are obviously uncorrelated. The large-scale investigation 
of the stronger lines provides marginal signals exceeding the $1 \sigma$ level 
at $\overline{\Delta v}=500$~{\kms} and $\overline{\Delta v}=7\,500$~{\kms} 
(Fig. \ref{tpcf_sl_10000_all}).

\begin{figure} 
\includegraphics[angle=90,scale=0.35]{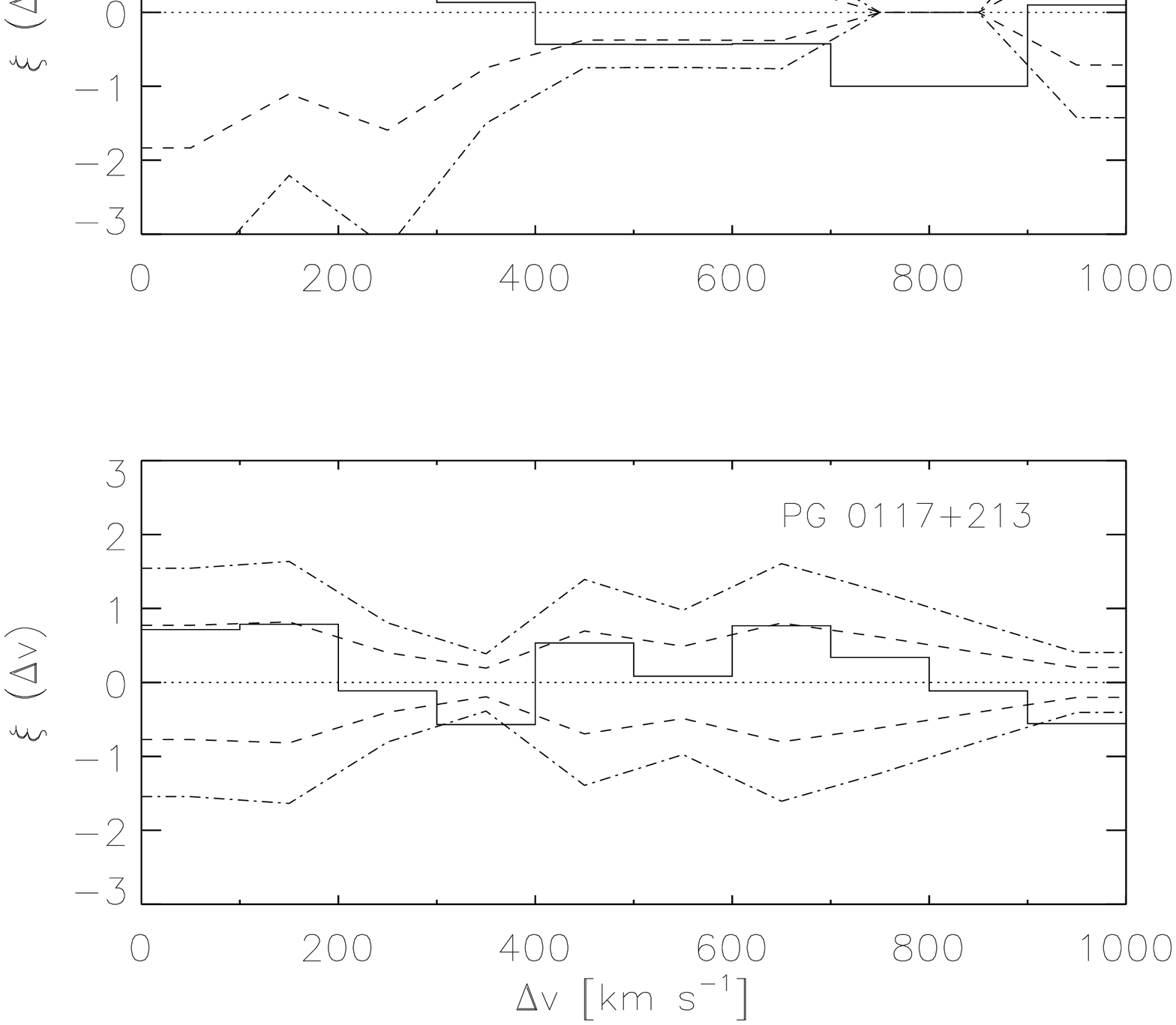} 
\caption[] 
{\label{tpcf_sl_1000_indivQSO} Two-point velocity correlation function 
for the stronger absorbers for individual quasar LOS, in 100 {\kms} bins.
Line symbols are the same as in Fig. \ref{tpcf_wl_1000_all}. 
When there are no pairs in a specific bin, formally $n_{\rm obs}=0$, $\xi=-1$,
and the $1\,\sigma$ and $2\,\sigma$ Poisson errors are 0, 
following error propagation as was done for all other bins.}
\end{figure}

The clustering of the high and low {\N} absorbers on short scales 
contradicts our own results presented in \cite{Janknecht2002}, where no 
clustering signal at all was found in the LOS to HE~0515-4414 alone, 
independent of the investigated {\N} ranges and velocity splittings.
Individual LOS might not offer a sufficient statistical basis 
to detect a weak clustering signal. This is supported 
by Fig.~\ref{tpcf_sl_1000_indivQSO} where we plot the correlation function 
$\xi$ for some individual LOS of our sample. Although the four LOS 
with the highest numbers of strong lines are considered ($ > 40$), 
occasionally an at most marginal excess above the $1 \sigma$ level 
can be seen for $\Delta v < 200$~{\kms}, but no value at all 
above the $2 \sigma$ level.

Most studies found significant clustering on short scales, even to some extent 
for the low {\N} lines \citep{Kim2001, Cristiani, Ulmer, Hu}.
 
Previous observations suggested a clustering signal increasing with
decreasing {\mbox{redshift}} \citep{Kim2001,Ulmer}. This is made plausible
by the fact that the structures in the universe develop by gravitational
pull over the course of time. In addition, at low {\z} $\gtrsim 1/3$ 
of all {\Lya} absorbers are associated with galaxies \citep{Lanzetta1995} 
which are known to cluster on these scales.

\begin{figure} 
\includegraphics[angle=90,scale=0.35]{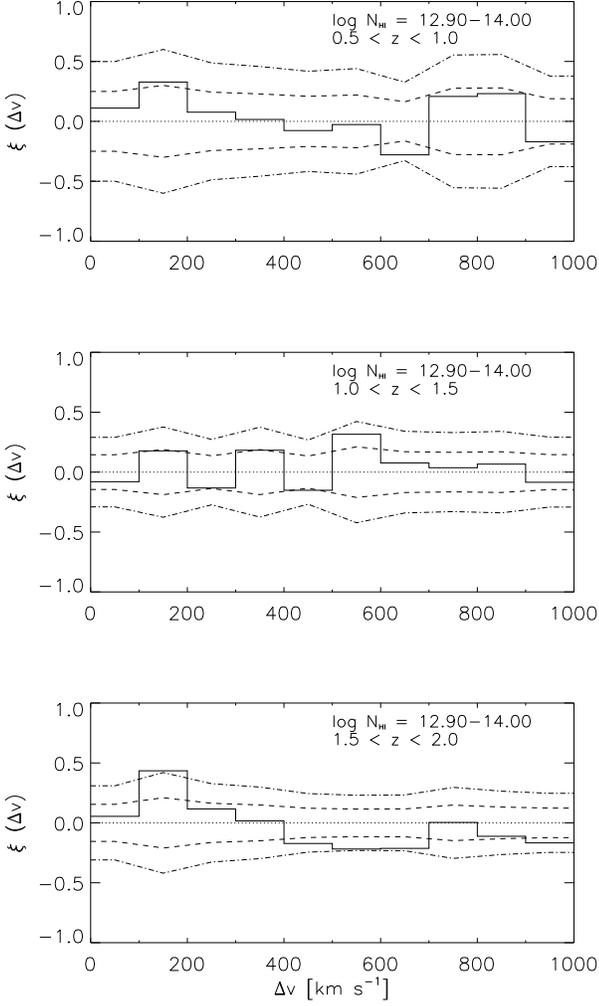} 
\caption[] 
{\label{tpcf_wl_1000_ev} Evolution of the two-point velocity correlation 
function for the weaker absorbers, in $100$~{\kms} bins, illustrated 
with three different $z$ intervals. Line symbols are the same 
as in Fig.~\ref{tpcf_wl_1000_all}.}  
\end{figure} 

\begin{figure} 
\includegraphics[angle=90,scale=0.35]{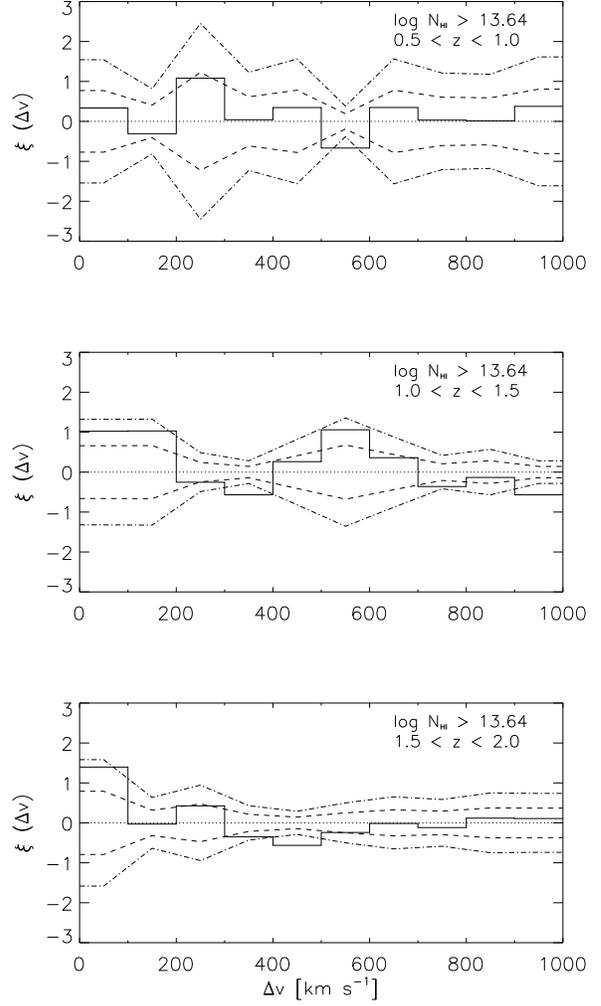}
\caption[] 
{\label{tpcf_sl_1000_ev} Evolution of the two-point velocity correlation 
function for the stronger absorbers, in $100$~{\kms} bins, illustrated 
with three different $z$ intervals. Line symbols are the same 
as in Fig.~\ref{tpcf_wl_1000_all}.}  
\end{figure} 

We also searched for a dependence of the correlation function on $z$. For
this purpose, we defined three $z$ intervals of the same width ($0.5-1.0$, 
$1.0-1.5$ and $1.5-2.0$). As described above we computed $\xi $ and 
estimated the $1\sigma$ and $2\sigma$ errors separately for each $z$
interval and for distances up to \mbox{$\Delta v_{\rm max}=1\,000$~{\kms}}
in $100$~{\kms} bins. The results are illustrated 
in Figs. \ref{tpcf_wl_1000_ev} and \ref{tpcf_sl_1000_ev}. 
For the weaker lines, the $\xi \approx 0$ value for $\Delta v<100$~{\kms} 
is again probably caused by blending effects, while for $\Delta v>200$~{\kms}
no clustering signal can be recognized. For $\overline{\Delta v}=150$~{\kms}
the strongest clustering is found in the highest {\z} range: 
$\xi =0.44 > 2\sigma$. At lower $z$, $\xi$ is roughly at the $1\sigma$ level 
($\xi=0.18$, $1 \sigma=0.19$ for $z=1.0-1.5$; $\xi=0.33$, $1 \sigma=0.30$ 
for $z=0.5-1.0$), i.e., for $z<1.5$, $\xi$ doesn't change relative to $\sigma$.

The high column density lines show a similar picture for 
$\overline{\Delta v}=50$~{\kms}: starting at a $\sim 2 \sigma$ level 
($\xi=1.40$, \mbox{$1 \sigma=0.79$}), the clustering degree decreases strongly 
with decreasing redshift. However, the opposite trend can be seen 
at $\overline{\Delta v}=150$~{\kms} where the signal increases from 
$\xi \approx 0$ to $\xi \approx 1.5\, \sigma$ when going from the highest 
to the middle $z$ interval. Besides, neither the high nor the low 
column density absorbers show a clustering signal lying clearly above 
the $2 \sigma$ level. 

In summary, no uniform evolutionary behaviour of $\xi$ can be established.
The rising correlation of the absorbers with decreasing redshift 
suggested by \cite{Kim2001} and \cite{Ulmer} cannot be confirmed.

\subsection{Evolution of the number density}
\label{nde}

As for the clustering properties, a different behaviour is also expected 
for the number density evolution of the stronger and weaker {\Lya} lines. 
We analyze this evolution using the two column density ranges from Subsection 
\ref{clust} separately.

In practice, the value for the differential number density 
$\frac{{\rm d}n}{{\rm d}z}$ can be deduced by choosing an interval width 
$\Delta z$ and counting the number of absorption lines $n$ 
in the individual intervals. When the line samples of several LOS 
which in general cover different $z$ regions are combined, 
it must be taken into account, 
of course, that a varying number of LOS $n_{\rm L}$ falls into a certain
{\z} interval. (Here, $n_{\rm L}$ can be estimated from Fig. \ref{LOS}).
Thus, we defined a density $\frac{{\rm d}n}{{\rm d}z}$ {\it per LOS}. 
Since the fluctuations between the LOS are large, only $z$ intervals 
with $n_{\rm L} > 1$ were considered. This is the case for 
$0.7 \leq z \leq 1.9$.  

We computed $\frac{{\rm d}n}{{\rm d}z}$ for the two sub-groups 
as a function of $z$. On the basis of the Levenberg Marquard algorithm, 
we fit the power law (\ref{dn/dz}) to the data points of
$\frac{{\rm d}n}{{\rm d}z}(z)$ and determined 
$\left(\frac{{\rm d}n}{{\rm d}z}\right)_{0}$ and $\gamma$ in this way.

\begin{figure}
\includegraphics[angle=90,scale=0.35]{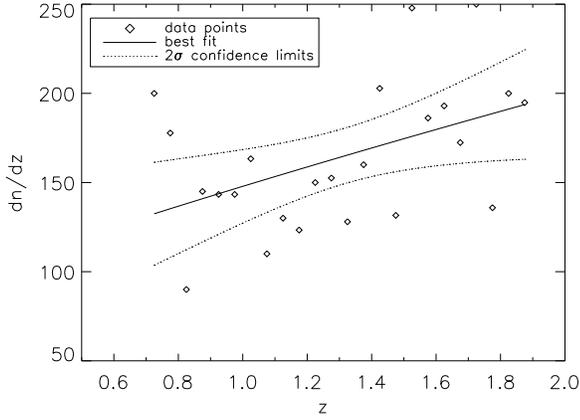}
\caption[]
{\label{weaklines} Number density evolution of the weak {\Lya} absorbers
($12.90 < {\rm log}\, N_{\rm \ion{H}{i}} < 14.00$). Given are the data points 
(binned with $\Delta z = 0.05$), the best fit and the $2 \sigma$ confidence 
limits.}
\end{figure}

\begin{figure}
\includegraphics[angle=90,scale=0.35]{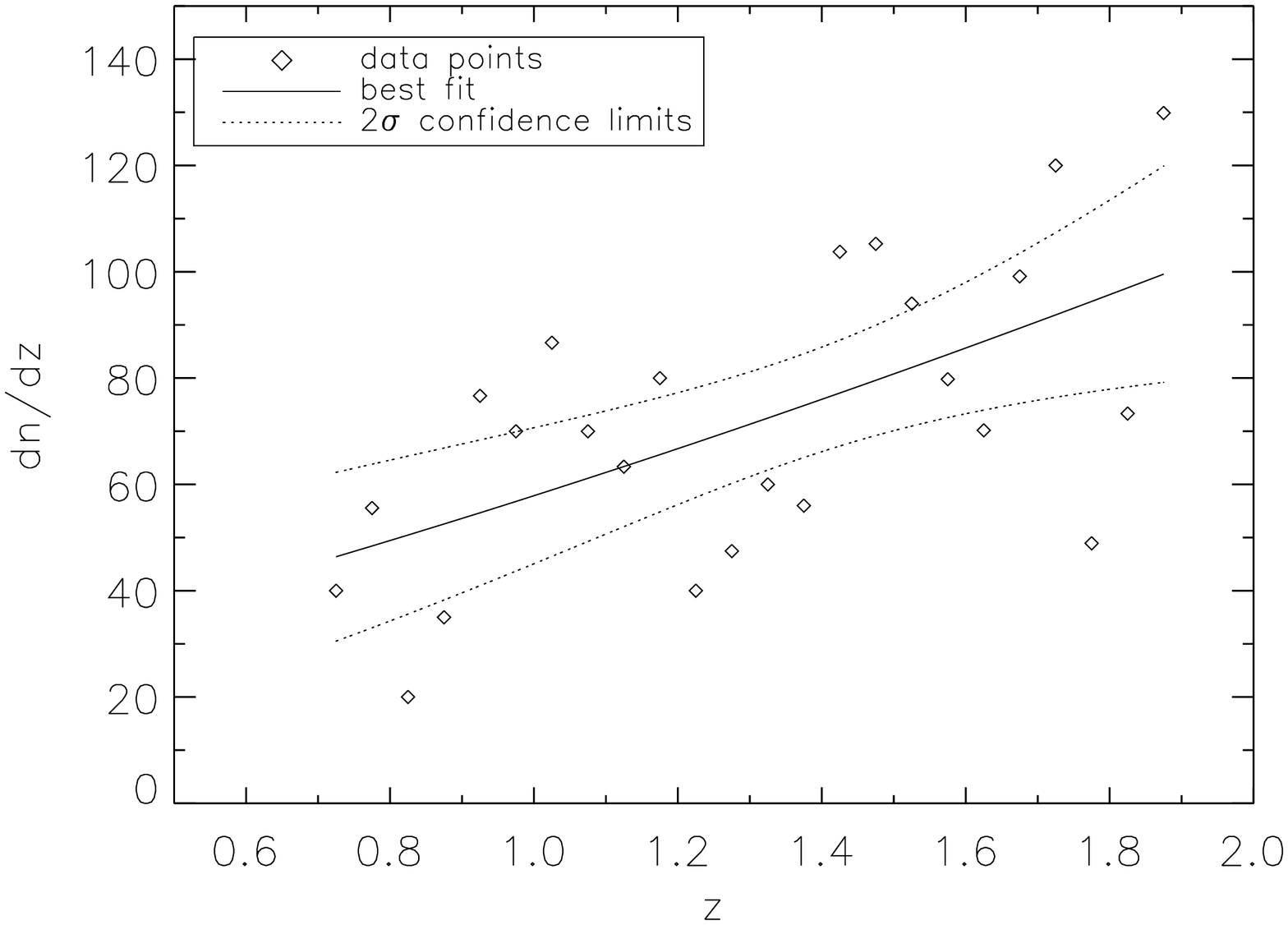}
\caption[]
{\label{stronglines} Same as in Fig. \ref{weaklines}, for the strong 
{\Lya} absorbers \mbox{(log~$N_{\rm \ion{H}{i}} > 13.64$)}.}
\end{figure}

Figs. \ref{weaklines} and \ref{stronglines} show the data points for the low
and high {\N} absorbers, respectively. In addition to the data points, 
the best fits to them with their $2\sigma$ confidence limits are presented.

For the weak absorbers, the number density decreases with decreasing redshift
with $\gamma=0.74\pm0.31$. The outliers in the interval $z=0.7-0.8$ might be
due to the small size of our sample in this region (see Figs. \ref{LOS} 
and \ref{Nz}). The stronger lines evolve faster with an exponent 
$\gamma=1.50\pm0.45$ which is only just consistent (within $1\sigma$) 
with the exponent of the weak lines.
 
\begin{figure}
\includegraphics[angle=90,scale=0.38]{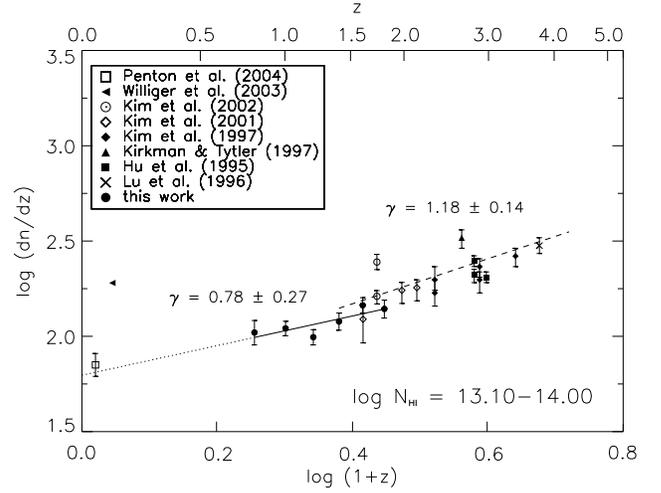}
\caption[]
{\label{nde_Lit_wl} Comparison of different studies 
of the number density evolution of the {\Lya} absorbers 
with $13.10 < {\rm log}\, N_{\rm \ion{H}{i}} < 14.00$. The {\Lya} lines 
of this work are binned in $\Delta z=0.2$ intervals. The solid line represents 
the best fit to the data points of this work, while the dotted line 
is an extrapolation of the fit for $z \rightarrow 0$. The dashed line 
shows the best fit to the data points in the higher redshift range, 
taken from \cite{Kim2002}.}
\end{figure}

\begin{figure}
\includegraphics[angle=90,scale=0.38]{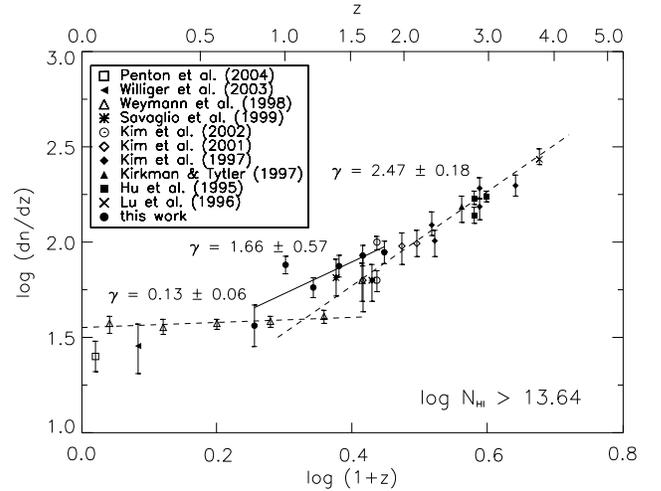}
\caption[]
{\label{nde_Lit_sl} Comparison of different studies 
of the number density evolution of the {\Lya} absorbers 
with log $N_{\rm \ion{H}{i}} > 13.64$. Details as in Fig. \ref{nde_Lit_wl}. 
The additionally given dashed line for log $(1+z)=0-0.4$ represents 
the best fit to the data points from \cite{Weymann}.
The data points from Penton et al. (2004), Savaglio et al. (1999) 
and Williger et al. (2003) apply to log $N_{\rm \ion{H}{i}} > 14.00$.}
\end{figure}

Fig. \ref{nde_Lit_wl} gives an extensive comparison of the results 
of this work for the weaker absorbers (here defined 
by \mbox{log~$N_{\rm \ion{H}{i}}=13.10-14.00$} to facilitate comparison with 
other results) with different contributions to the number density evolution 
from the literature measured in varying $z$ regions. Fig.~\ref{nde_Lit_sl} 
does the same for the stronger absorbers. In the double-logarithmic 
presentations the symbols correspond to the number densities 
measured in different studies. The {\Lya} lines of this work are binned 
in intervals of $\Delta z=0.2$ in order to have interval sizes comparable 
with the other investigations. The data points of this study are shown 
with filled circles; the error bars result from a $1\sigma $ poisson error 
of $n$.

Regarding the evolution of the weaker absorbers (Fig.~\ref{nde_Lit_wl}), it
becomes clear that the LOS analyzed here fall into a $z$ region which has
not yet been examined so far, due to a lack of high-resolution UV data. The
best fit to the data points of this work gives $\gamma =0.78\pm 0.27$.

The decrease of the line density with decreasing redshift is obviously
decelerated in our investigated $z$ interval, as is revealed from a
comparison with higher redshifts: for $z>1.5$, \cite{Kim2002} derived 
$\gamma =1.18\pm 0.14$ from a best fit to the data points of different
studies (presented as dashed line in Fig.~\ref{nde_Lit_wl}). Excluding two
LOS, they found $\gamma =1.42\pm 0.16$ which is no longer in agreement with
the result found here.

Interestingly, an extrapolation of our fit line for $z\rightarrow 0$ is
consistent within $1\sigma $ with the study of \cite{Penton2004} for the
local universe (see dotted line in Fig.~\ref{nde_Lit_wl}), while an
extrapolation of the \cite{Kim2002} fit line misses the \cite{Penton2004}
data point by $\approx 2\sigma$. (\cite{Williger2003} examine only a single
LOS). We interpret this as further evidence for a changing evolution of the
lower {\N} lines in the interval $z=1-2$.

Furthermore, the comparison with the investigations at higher and at lower $z$ 
suggests a continuous and moderate transition to the present, 
rather than a sharp break.

As already derived from Figs. \ref{weaklines} and \ref{stronglines}, the
high {\N} absorbers show a steeper gradient in the evolution diagram 
with $\gamma =1.66\pm 0.57$ (Fig.~\ref{nde_Lit_sl}). A remarkable data point 
occupies the interval $z=0.9-1.1$. Without this outlier the evolution index 
would be much higher and its error much lower ($\gamma =2.11\pm 0.21$). 
A detailed analysis of this {\z} range which is covered by six of the nine LOS 
(see Fig.~\ref{LOS}) shows that the high line density is not caused 
by a fluctuation in an individual LOS, but that the $\frac{{\rm d}n}{{\rm d}z}$
lies clearly above the level of the adjacent intervals in nearly all six LOS. 
Obviously we detect a coincidental line accumulation within this interval.

Just as for the weaker lines, our $\gamma $ is lower compared with the
results of investigations at higher $z$: \cite{Kim2002} found 
$\gamma=2.47\pm 0.18$ by fitting a straight line to the data points of varying
studies for $z>1.5$ which is inconsistent within 1$\sigma $ with the value
computed here. In addition, the number density continuously decreases in the
analyzed range $z=0.7-1.9$ (apart from the outlier) without showing the
frequently postulated break in the evolution at $z=1.5-1.7$ 
\citep{Weymann, Dobrzycki, Impey1996}. Possibly this slow-down 
in the evolution, which was also successfully simulated by theoreticians 
\citep{Theuns1998, Dave1999}, takes place in a later phase ($z<0.7$) 
beyond the redshift region examined here. An indication for this might be 
that at lower $z$ the thinning out of the absorbers obviously continues 
to decelerate: in contrast to the weaker absorbers, an extrapolation 
of our fit line for $z\rightarrow 0$ would lead to a distinctly too low 
absorber density at $z=0$, compared with \cite{Penton2004} and \cite{Weymann} 
(see Fig.~\ref{nde_Lit_sl}).

If we fit the data points of \cite{Weymann} with a straight line (shown in
Fig.~\ref{nde_Lit_sl}; $\gamma=0.13\pm0.06$), it does not intersect our fit
line before $z=0.6$. Also, leaving out our outlier leads to a point of
intersection at $z=0.8$. Thus, in the case that the break in the number
density evolution took place, the redshift value at which it occurred has to
be corrected downwards.

However, it must be noted that \cite{Weymann} used the equivalent width
$W_{\lambda}$ instead of the column density as the observable. 
The conversion formula

\begin{equation}
\label{Aequi}
W_{\lambda,\,{\rm rest}}\, [{\rm \AA}]= 8.85\cdot10^{-21}\,\,
\big(\lambda_{\rm rest}\, [{\rm \AA}]\big)^{2}\, f\,\, N\, 
\left[{\rm cm}^{-2}\right]
\end{equation}
($f$ oscillator strength) only applies to unsaturated absorption lines 
and depends on the Doppler parameter. Occasionally, 
\mbox{$b=25\,{\rm km\,s}^{-1}$} is assumed in the literature 
\citep{Penton2000, Penton2004}. Then the lower limit for the equivalent width 
\mbox{$W_{\lambda,\,{\rm rest}}=240$ m\AA\,} used by \cite{Weymann} 
corresponds to a minimum column density log~$N_{\rm \ion{H}{i},\, min}=14.00$. 
Applying this higher cut-off to the evolution diagram of the stronger absorbers
would shift our data points in Fig.~\ref{nde_Lit_sl} significantly downwards. 
As a result of this, the intersection of our fit line with the one 
to the \cite{Weymann} data points (and therefore the evolution break 
estimated from it) had to be corrected to a higher redshift. 
For this reason, the uncertainty in the 
\mbox{$W_{\lambda}$-$N_{\rm \ion{H}{i}}$} conversion makes 
the exact localization of the slow-down difficult to determine.

Nevertheless, newer studies of \cite{Kim2001}, \cite{Kim2002} (whose fit
line hits that of the \cite{Weymann} data points at $z=1.1$; 
see Fig.~\ref{nde_Lit_sl}) or \cite{Penton2004} also favoured the break 
in the evolution taking place at no higher redshift than $z=1.0-1.2$. 
Supported by these results, we conclude that a transition in the evolution 
of the higher column density lines indeed took place, but much later 
than assumed before. In simulations the sudden slow-down in the evolution 
results directly from the decline of the UV background at $z=2$ 
\citep{Dave1999}. However, the QSO dominated UV background of \cite{Haardt1996}
is usually assumed in these calculations, and thus the contribution 
of the galaxies is possibly underestimated. With the latter as main source 
of the UV radiation field for $z<2$, the radiation field would decrease 
more slowly and the break would naturally be corrected to a lower $z$.

\vspace{1cm}

All results for the evolution of the number density of {\Lya} absorbers 
in this subsection have to be considered in view of the high scatter 
of the data points. This scatter could not be significantly reduced
compared with our study of the single LOS to HE~0515-4414 
\citep{Janknecht2002}, though we enhanced the examined total redshift path
length $\Delta z$ by roughly a factor 6. For all nine observed QSOs we
obtained high-resolution spectra ($R\geq 30\,000$) where the {\Lya} forest 
is completely resolved (because $b_{\rm \ion{H}{i}} > 10$ {\kms}).
Therefore, we suppose that the variations from LOS to LOS are caused 
{\it physically} (rather than methodically or statistically), 
reflecting a strongly inhomogenous appearance of the {\Lya} forest. 
Indeed, \cite{Kim2002} and \cite{Tytler2004} (for $z<2.5$ and for $z=1.9$, 
respectively) and \cite{Impey1999} for the local universe ($z<0.2$) 
found hints for a strong cosmic variance too.

The reason for the large variations of $\frac{{\rm d}n}{{\rm d}z}$
between individual LOS falling into the same redshift ranges is probably the
advanced stage of structure formation in the universe at $z=2$. 
The structures are gravitationally evolved: over- and underdensities 
have been formed. In individual LOS this can be recognized in the form 
of strongly varying number densities, in particular of the high column density
absorbers. The structure formation obviously starts to dominate over the
Hubble expansion, which causes the general decline of the number density.

\subsection{Effective optical depth at $z \approx 2$} 
\label{OptTiefe}

We examined the effective optical depth $\tau_{\rm eff}$ 
and the average normalized flux $\left<F_{\rm norm}\right>$,
which are related to each other by the equation

\begin{equation} 
\label{Fnorm_Def} 
\tau_{\rm eff} = - {\rm ln} \left <\frac{F}{F_{\rm cont}} \right> 
\equiv - {\rm ln} \left <F_{\rm norm} \right>,
\end{equation}   
where $F_{\rm cont}$ is the continuum flux. We calculated $\tau_{\rm eff}$ and 
$\left <F_{\rm norm} \right>$ for the maximum redshift range of our LOS,
\mbox{$z=1.81-1.91$}. 
The LOS to three of our quasars fall into the chosen interval:  
those to HE~2225-2258, HE~0429-4901 and HS~0747+4259 (see Fig.~\ref{LOS}).
Therefore, we used the flux distributions of the corresponding spectral parts 
for the calculation of $\left<F_{\rm norm}\right>$. Because we were interested
in the H{\sc i} absorption of the low density IGM, we defined wavelength 
regions around the metal lines occuring in the regarded $z$ interval 
and eliminated these regions from the three spectra. The remaining 
flux distribution $F(\lambda)$ was divided by the respective continuum flux 
$F_{\rm cont}(\lambda)$ which is an output parameter of CANDALF. 
For the resulting normalized flux averaged over the wavelength positions 
of all three spectral parts we derived

\begin{equation} 
\label{Fnorm_Res} 
\left<F_{\rm norm}\right > = 0.895\pm0.020
\end{equation} 
for $<z> \approx 1.86$, i.e. 10.5\% of the radiative flux is absorbed
by the diffuse intergalactic medium at this redshift.
The error of $\left<F_{\rm norm}\right>$ follows from Tytler et al. (2004),
eq. (30): since we have only three LOS, the total error 
of our average normalized flux is completely dominated 
by the large scale structure and will probably be $>0.02$.
With (\ref{Fnorm_Def}) and (\ref{Fnorm_Res}) we compute

\begin{equation} 
\label{taueff_Res} 
\tau_{\rm eff} = 0.111\pm0.020
\end{equation} 
for the effective optical depth. 

\cite{Kirkman2005} investigated the H{\sc i} opacity of the IGM at 
$1.6 < z < 3.2$. At z=1.86, the scaling in their Fig.~5 gave
$\left<F_{\rm norm}\right >=0.889$ with an error of roughly 0.01.
Thus, our result agrees with theirs.


\section{Conclusions}

We have analyzed the combined sample of all {\Lya} absorption lines 
which were detected in the spectra of nine quasars 
over a total redshift path length $\Delta\, z=5.176$ within the range 
\mbox{$z=0.5-1.9$}. We classified 1325 absorption features as {\Lya} lines. 
Examining the distributions of their fit parameters {\logn}, $b$ and $z$, 
we find the following:

\begin{enumerate} 

\item{The {\N} distribution might be complete down to 
log~$N_{\rm \ion{H}{i}}=12.90$. It can be approximated over roughly 
three orders of magnitude (log~$N_{\rm \ion{H}{i}}=12.90-15.70$) 
by a simple power law. The slope $\beta=1.60\pm0.03$ is consistent 
with other studies in comparable redshift intervals. We do not see any change 
in $\beta$ with $z$.}  

\item{The distribution of the Doppler widths has the expected form of nearly 
Gaussian with an additional expanded tail to high $b$ and an average value 
$\overline{b}=(34\pm22)\, {\rm km\, s}^{-1}$. We cannot detect any evolution
of the Doppler parameter distribution with $z$ in the examined {\z} phase.}

\item{The weaker (log~$N_{\rm \ion{H}{i}}=12.90-14.00$) as well as 
the stronger {\Lya} lines (log~$N_{\rm \ion{H}{i}} > 13.64$) show 
marginal clustering with a $2\sigma $ significance 
on short velocity intervals ($\Delta v < 200$~{\kms} and 
\mbox{$\Delta v < 100$~{\kms}}, respectively).
However, the clustering signal is too weak to be seen in individual LOS.
We do not see any change with column density or redshift.}
 
\item{Using the customary power law 
$\frac{{\rm d}n}{{\rm d}z} \propto (1+z)^{\gamma}$, the evolution 
of the number density of the weaker {\Lya} lines declines with 
$\gamma=0.74\pm0.31$ with decreasing $z$ within the range $z=0.7-1.9$.
From comparisons with investigations of other authors at higher and 
lower redshifts we conclude that the decrease of the weaker lines 
is decelerated in the phase \mbox{$z=1-2$}, turning into a nearly 
flat evolution for $z \rightarrow 0$, without showing any hint for
a sharp break in the evolution.}

\item{The number density of the stronger absorbers also decreases 
with decreasing $z$ ($\gamma=1.50\pm0.45$ for $z=0.7-1.9$), 
faster than for the weaker lines, though the trends
are consistent with each other within $1\sigma$.}

\item{The decline of the line density of the stronger absorbers is also
decelerated compared with higher {\z}s \mbox{($z=1.5-4.0$)}. The break 
in the evolution predicted for $z=1.5-1.7$ cannot be detected, however.  
The deduced $z$ dependence of the number density as well as comparisons 
with other studies for the local universe rather suggest a later slow-down 
($z < 0.7$), followed by an approximately flat evolution.
The precise $z$ value of the break in the evolution probably has to be 
corrected somewhat upwards due to the uncertain conversion between 
equivalent width (used by \cite{Weymann} as observable) and column density.}

\item{Though having analyzed nine high-resolution ($R \geq 30\,000$) 
quasar spectra altogether, the scatter of the data points 
in the diagrams of the number density evolution is large. 
As Tytler et al. (2004) showed, the strong variations between the LOS 
are consistent with the effects of normal structure formation.}

\item{An analysis of the flux distribution of the three quasars with LOS 
in the upper {\z} range ($z=1.81-1.91$) gives 
$\left<F_{\rm norm}\right > = 0.895\pm0.020$ for the average normalized flux 
and $\tau_{\rm eff} = 0.111\pm0.020$ for the effective optical depth, 
in good agreement with the values from Kirkman et al. (2004).}

\end{enumerate}

\begin{acknowledgements}

This research has been supported by the Verbundforschung 
of the BMBF/DLR under Grant No.~50~OR~9911~1.
SL was supported by the Chilean {\sl Centro de Astrof\'{i}sica} FONDAP
No. 15010003, and by FONDECYT grant N$^{\rm o} 1030491$. DT was supported 
in part by NSF grant AST-0507717, and by NASA grants HST-AR-10288 
and HST-AR-10688.

We thank the anonymous referee for his helpful report. We also thank 
Mr. Carl Zeisse who edited the text and gave helpful comments.

\end{acknowledgements}

\end{document}